\documentclass{jaa}
\usepackage{natbib} 
\usepackage{hyperref}
\usepackage{amsmath}
\usepackage{amstext}
\usepackage{amssymb}
\usepackage{url}
\usepackage{txfonts}
\usepackage{balance}
\usepackage{textcase}
\usepackage{float}
\usepackage{rotating,longtable}
\usepackage{xspace}
\usepackage{hyperref}
\usepackage{graphicx}
\usepackage{xcolor}
\usepackage{graphicx}
\usepackage{caption}
\usepackage{subcaption}
\usepackage{lipsum} 

\usepackage{lineno}

\usepackage[T1]{fontenc}
\usepackage{ae,aecompl}

\begin{document}\sloppy

\title{Kronberger 55: A Candidate for End-dominated Collapse Scenario}

%

\author{Aayushi Verma\textsuperscript{1,*}, Saurabh Sharma\textsuperscript{1}, Lokesh Dewangan\textsuperscript{2}, Rakesh Pandey\textsuperscript{2}, Tapas Baug\textsuperscript{3}, Devendra
K. Ojha\textsuperscript{4}, Arpan Ghosh\textsuperscript{1} and Harmeen Kaur\textsuperscript{5}}
\affilOne{\textsuperscript{1}Aryabhatta Research Institite of observational sciencES (ARIES), Manora Peak, Nainital-263001, India\\}
\affilTwo{\textsuperscript{2}Physical Research Laboratory, Navrangpura, Ahmedabad—380 009, India\\}
\affilThree{\textsuperscript{3}Satyendra Nath Bose National Centre for Basic Sciences (SNBNCBS), Block-JD, Sector-III, Salt Lake,
Kolkata-700 106, India\\}
\affilFour{\textsuperscript{4}Tata Institute of Fundamental Research (TIFR), Homi Bhabha Road, Colaba, Mumbai - 400 005, India\\}
\affilFive{\textsuperscript{5}Center of Advanced Study, Department of Physics DSB Campus, Kumaun University Nainital, 263002,
India\\}

\twocolumn[{

\maketitle

\corres{aayushiverma@aries.res.in}

%
\msinfo{23 November 2022}{12 January 2023}{13 January 2023}

\begin{abstract}
Using optical photometric observations from 1.3m Devasthal Fast Optical Telescope and deep near-infrared (NIR) photometric observations from TANSPEC mounted on 3.6m Devasthal Optical Telescope, along with the multi-wavelength archival data, we present our study of open cluster Kronberger 55 to understand the star formation scenario in the region. The distance, extinction and age of the cluster Kronberger 55 are estimated as $\sim$3.5 kpc, $E(B-V)\sim$1.0 mag and $\lesssim$5 Myr, respectively.
We identified Young Stellar Objects (YSOs) based on their excess infrared (IR) emission using the two-color diagrams (TCDs). The mid-infrared (MIR) images reveal the presence of extended structure of dust and gas emission along with the outflow activities in the region with two peaks, one at the location of cluster Kronberger 55 and another at $5^{\prime}.35$ southwards to it. The association of radio continuum emission with the southern peak, hints towards the formation of massive star/s. The \emph{Herschel} sub-millimeter maps reveal the presence of two clumps connected with a filamentary strcuture in this region, and such configuration is also evident in the $^{12}$CO$(1-0)$ emission map. Our study suggests that this region might be a  hub-filament system undergoing star formation due to the `end-dominated collapse scenario'. 
\end{abstract}

\keywords{Interstellar medium, H $\textsc{ii}$ regions, Open star clusters, Star formation, Star forming regions.\\}}]


\doinum{12.3456/s78910-011-012-3}
\artcitid{\#\#\#\#}
\volnum{000}
\year{0000}
\pgrange{1--}
\setcounter{page}{1}
\lp{1}

\section{Introduction}\label{sec:intro}

The recent advancement in the IR and submillimeter (submm) data provided by space-based telescope, such as \emph{Herschel} and \emph{Spitzer}, has taken us towards commendable observational progress to study star formation.   
Recently, using high-resolution images of \emph{Spitzer} and \emph{Herschel}, many authors have shown that the cold interstellar medium (ISM) have filamentary structures around the birthplace of stars (for e.g., \citealt{Churchwell_2006}, \citeyear{Churchwell_2007}; \citealt{2010A&A...518L.102A}).  
The role of the filaments in the star formation processes is yet to be known. The zones where filaments either merge or collide with each other, are considered as appropriate environment for the formation of star clusters (\citealt{2010A&A...518L.102A}, \citeyear{2014prpl.conf...27A}; \citealt{Nakamura_2012}, \citeyear{Nakamura_2014}; \citealt{2013MNRAS.428.3425H}; \citealt{2017ApJ...834...22D}, \citeyear{2019ApJ...884...84D}; \citealt{2022ApJ...930..169B}; \citealt{2022ApJ...934....2M}). These zones are known as `hub' and such system is known as 
`hub-filamentary system' (HFS).
Along with HFSs, isolated filaments also exist in our galaxy which are associated with star-forming activities. The end-dominated collapse (EDC; or edge-collapse) process is associated with such isolated filaments where two clumps/cores are expected to be produced at ends of the filament through high gas acceleration \citep{1983A&A...119..109B, Pon_2012,2022arXiv220307002H}. This scenario is of interest for many recent investigation which is possible after the availability of high quality \emph{Herschel} data along with velocity information from the CO maps  (\citealt{Goicoechea_2015}; \citealt{2016ApJ...819...66D}, \citeyear{2022ApJ...925...41D}; \citealt{2020A&A...641A..24W}). 
But, till now, such process has been reported in a few star-forming sites (for e.g., \citealt{Dewangan_2017}, \citeyear{Dewangan_2019}; \citealt{Bhadari_2020}, \citeyear{2022ApJ...930..169B}). 

\begin{figure*}
    \centering
    \includegraphics[width=0.65\textwidth]{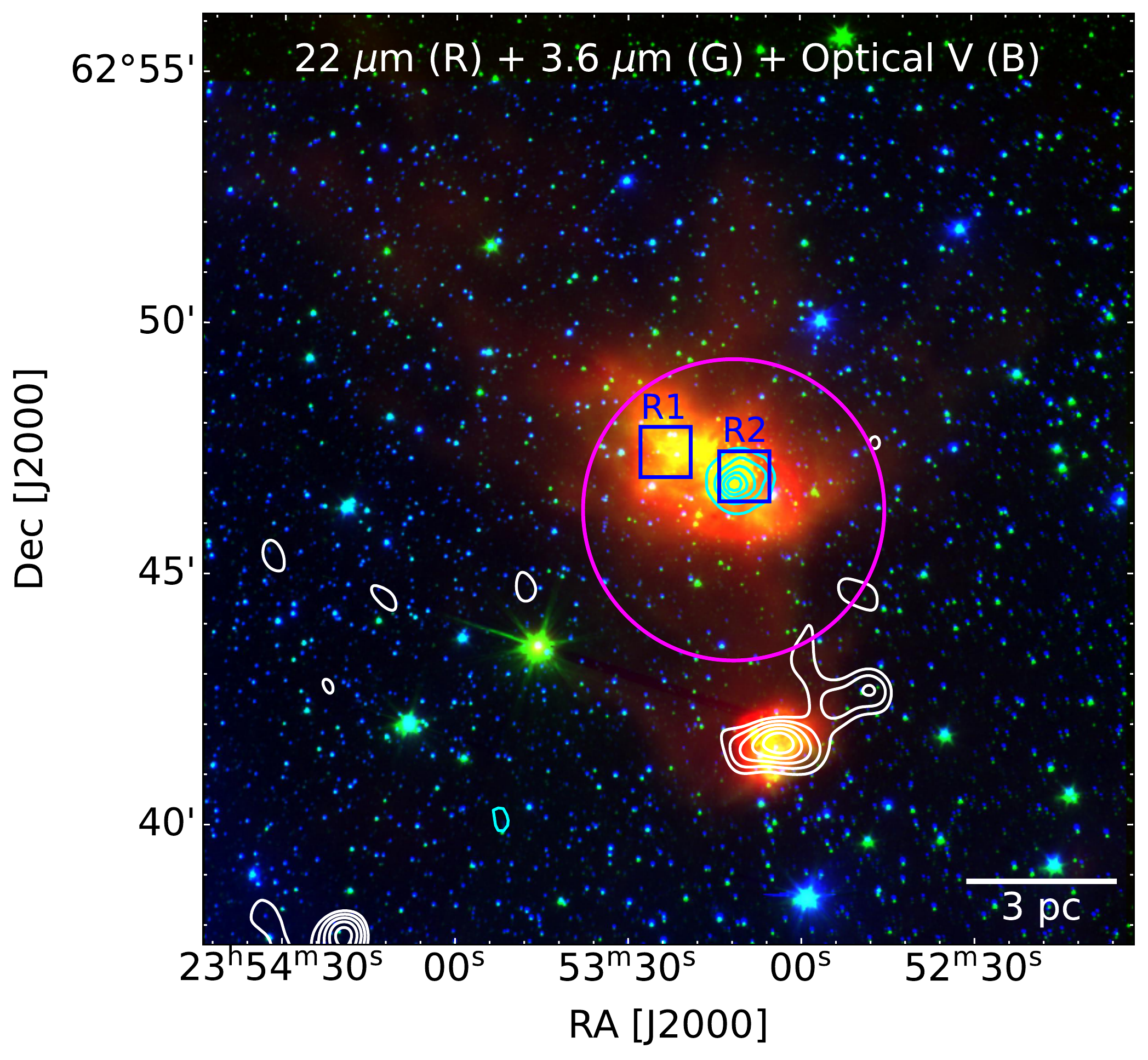}
    \caption{Color-composite image (Red: \emph{WISE} 22 $\mu$m; Green: \emph{Spitzer} 3.6 $\mu$m and Blue: Optical V band (taken using 1.3m DFOT) of $18^\prime.5 \times18^\prime.5$ region (field-of-view of the optical observations) overlaid with the isodensity contours generated from the 2MASS catalog (cyan) and NVSS 1.4 GHz radio continuum contours (white). The lowest level for the isodensity contours is 4$\sigma$ above the mean stellar density (i.e., 3 stars arcmin$^{-2}$) with a step size of 1.25 stars arcmin$^{-2}$ whereas for the generation of radio map, it is 1.2 mJy with a step size of 0.186 mJy. Blue squares represent the region selected to be observed using TANSPEC and are termed as R1 and R2 whereas magenta circle represents active region of the cluster.}
    \label{fig:intro_rgb}
\end{figure*}

\begin{figure*}
    \centering
    \includegraphics[width=0.3\textwidth]{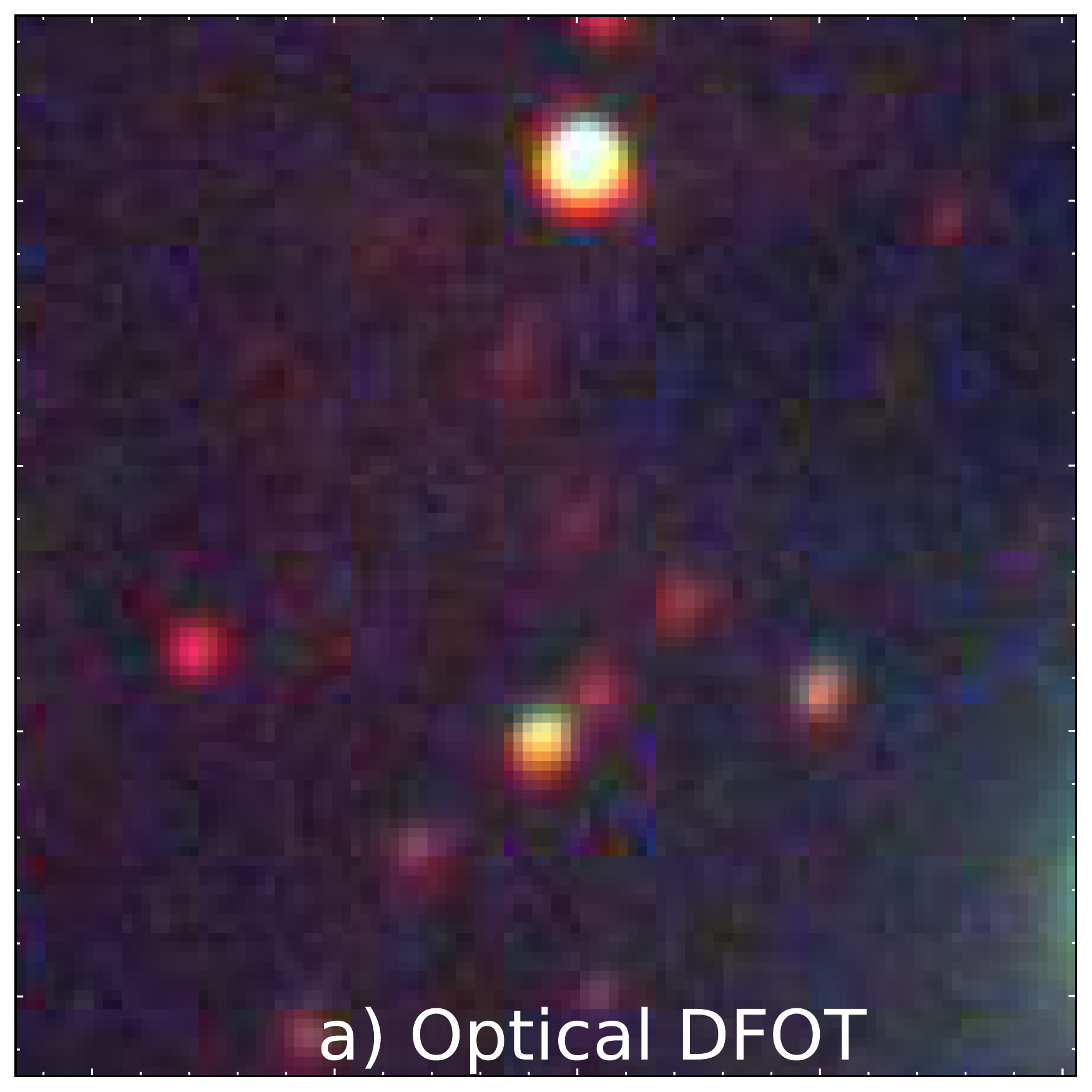}
    \includegraphics[width=0.3\textwidth]{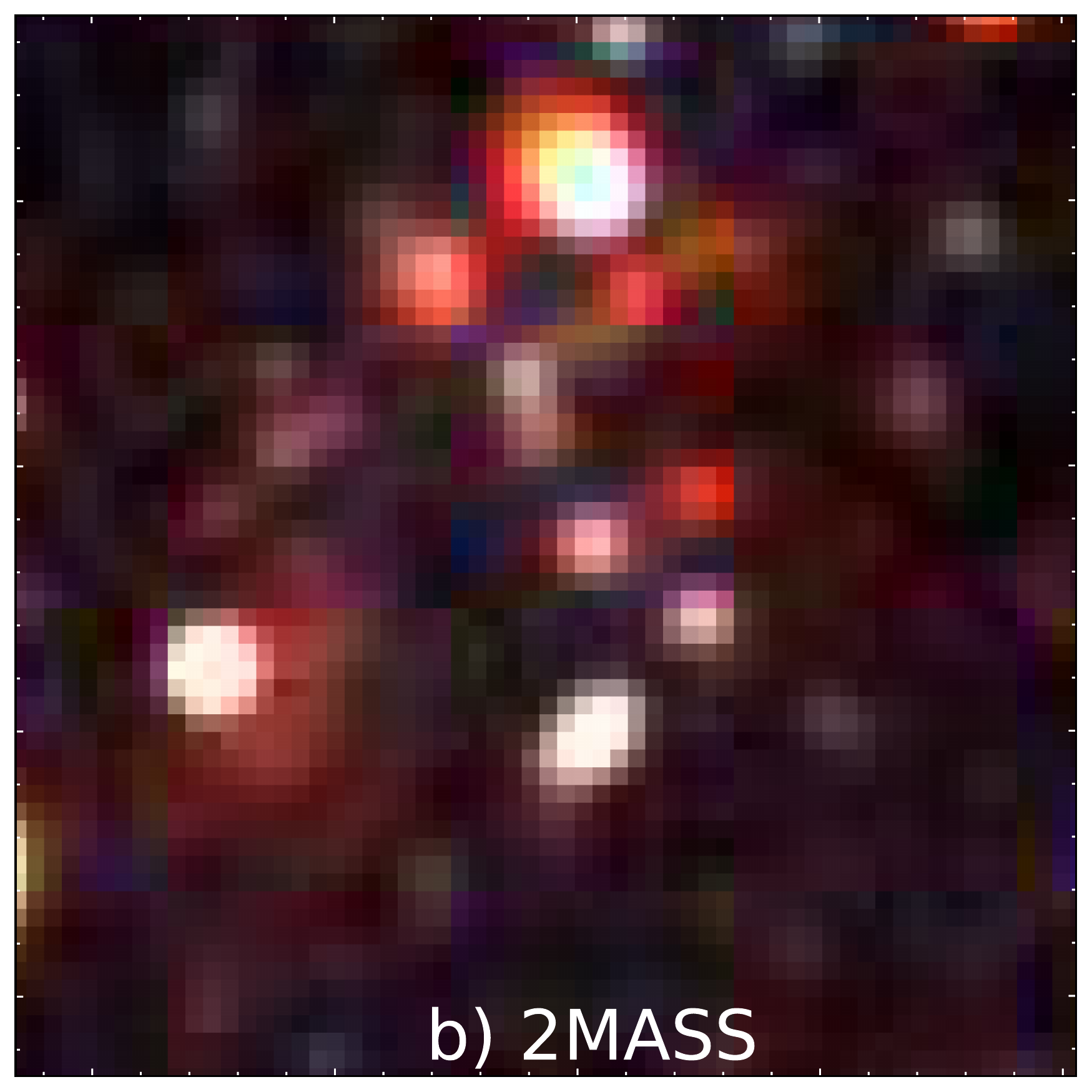}
    \includegraphics[width=0.3\textwidth]{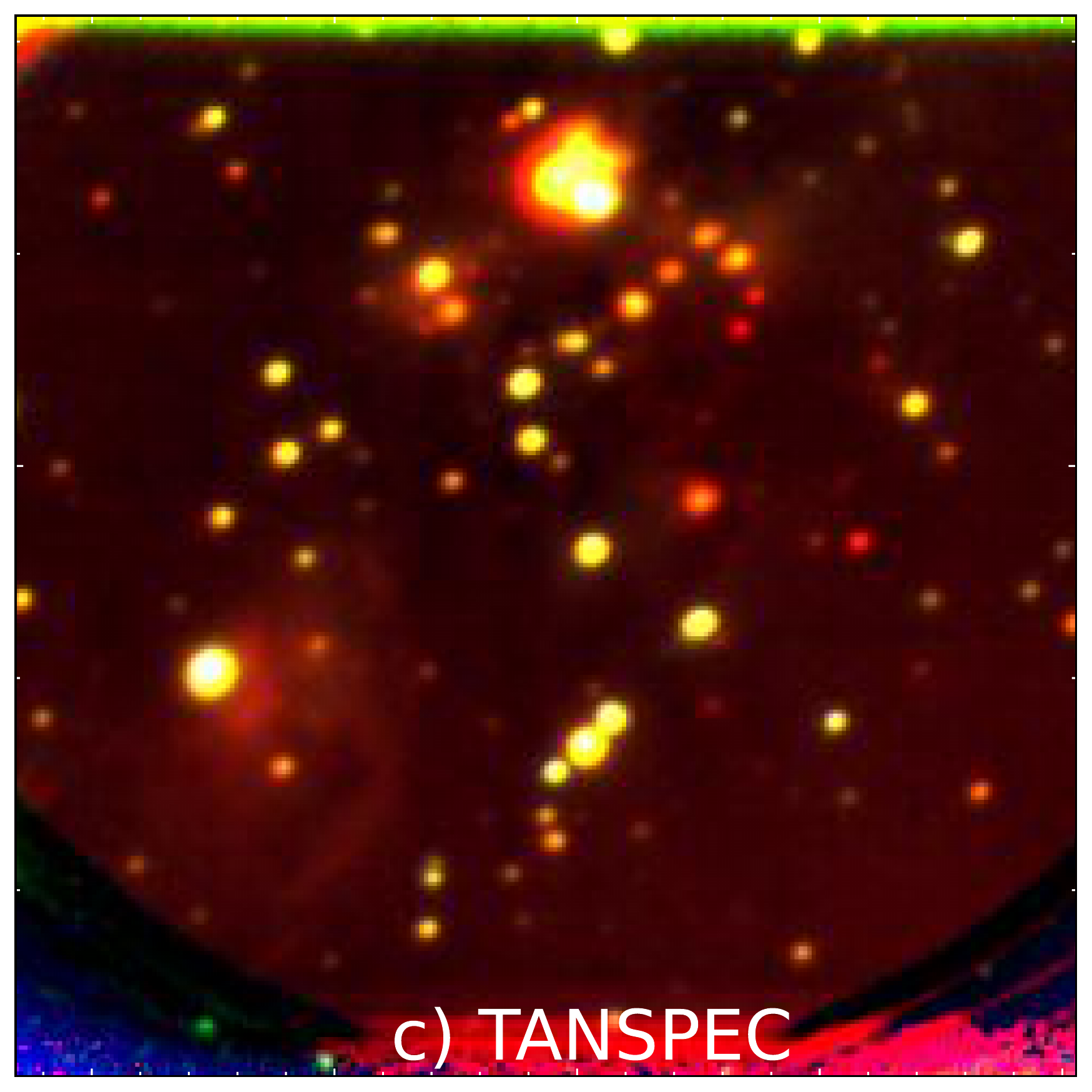}
    \caption{The color-composite images for the zoomed-in view of R2. The left panel is the color-composite image (Red: $I_c$; Green: $V$ and Blue: $B$ band) of DFOT and the middle and right panels are color-composite images (Red: $K_s$; Green: $H$ and Blue: $J$ band) of 2MASS and TANSPEC respectively.}
    \label{fig:intro_rgb_comparison}
\end{figure*}

For the present study, we have selected a less explored open cluster Kronberger 55 ($\alpha=23^h53^m23^s$ and $\delta=+62^{\circ}46^{'}54^{''}$) \citep[][]{2002A&A...389..871D, 2006A&A...447..921K}, which seems to be embedded in a filamentary structures with signatures of star formation activities (cf. Figure \ref{fig:intro_rgb} and Figure \ref{fig:intro_rgb_comparison}).
\citet{2002A&A...389..871D} 
reported the distance, reddening and age  of this cluster as 1.26 kpc, $E(B-V)$=1.13 mag and 400 Myr, respectively. \citet{2009Ap&SS.323..383T} reported its age as 400 $\pm$ 20 Myr, its distance as 1.20 $\pm$ 0.06 kpc and interstellar reddening ($E_{B-V}$) as 1.13 $\pm$ 0.11 mag by using the 2MASS data. Recently, \citet{2016A&A...585A.101K} did a global survey of star clusters based on the uniform cluster membership estimated using the 2MAst (i.e. 2MASS with astrometry) catalogue, a merger of the Two Micron all Sky Survey (2MASS) data and PPMXL (a catalog of proper motions, positions, 2MASS- and optical photometry of 900 million stars and galaxies which aims to be complete down to $V=20$.). The estimated distance, reddening and the age of this cluster was reported as 1.99 kpc, $E_{B-V}$=0.96 mag and 24 Myr. However, above estimations are affected by larger uncertainties by the shallow photometric completeness ($V\sim$20 mag, $J\sim$15.3 mag) and high extinction around the Kronberger 55 cluster. 
A deeper imaging data can help us to characterize the faint low-mass young stars still embedded in the molecular cloud.
Hence, we present a detailed analysis of this cluster and the surrounding region by using deep optical/near-infrared (NIR) data observed through the 1.3m/3.6m telescopes at Devasthal, Nainital, India \citep{2018BSRSL..87...29K} along with MIR/radio maps from various data archives.

The structure of the paper is as follows: Section \ref{sec:observation} describes the observed data and the followed data reduction procedures along with the used archival data. Section \ref{sec:result} describes the estimation of the basic parameters this cluster along with the surrounding physical environment. In Section \ref{disc}, we have discussed the star formation scenario of this cluster and in Section \ref{sec:conclusion} we have concluded our study.

\section{Observation and Data Reduction}\label{sec:observation}

\subsection{Optical Photometric Observation and Data Reduction}
Optical observation of the region was done on  October 13, 2021 with 1.3m Devasthal Fast Optical Telescope (DFOT), Nainital \citep{2012ASInC...4..173S}, in broadband $UBVI_c$ filters using $2K \times 2K$ CCD (plate scale = 0.54 arcsec/pixel) camera which has a field of view (FOV) of $18^\prime.5 \times18^\prime.5$. 
Few flat and bias frames were also taken along with the science frames.
The image processing and photometric measurements were done through IRAF reduction packages along with DAOPHOT-II routines (outlined in \citealt{2020ApJ...891...81P}). The instrumental magnitudes of the stellar objects were calibrated by standard stars located in SA95 \citep{1992AJ....104..340L}. The procedure outlined by \citet{1992ASPC...25..297S} was followed for the same. Following are the calibration equations obtained by least-square linear regression:

\begin{equation}
    \begin{split}
        u& = U + (4.691 \pm 0.006)\\
        &- (0.093 \pm 0.005)(U-B) + (0.611 \pm 0.006)X_U,
    \end{split}
\end{equation}

\begin{equation}
    \begin{split}
        b& = B + (2.900 \pm 0.005)\\
        &- (0.126 \pm 0.003)(B-V) + (0.284 \pm 0.009)X_B,
    \end{split}
\end{equation}

\begin{equation}
    \begin{split}
        v& = V + (2.290 \pm 0.003)\\
        &+ (0.101 \pm 0.002)(V-I_c) + (0.130 \pm 0.005)X_V,
    \end{split}
\end{equation}
\\and\\
\begin{equation}
    \begin{split}
        i_c& = I_c + (2.516 \pm 0.007)\\
        &- (0.061 \pm 0.004)(V-I_c) + (0.058 \pm 0.014)X_I.
    \end{split}
\end{equation}

\begin{figure}[!ht]
    \centering
    \includegraphics[width=0.48\textwidth]{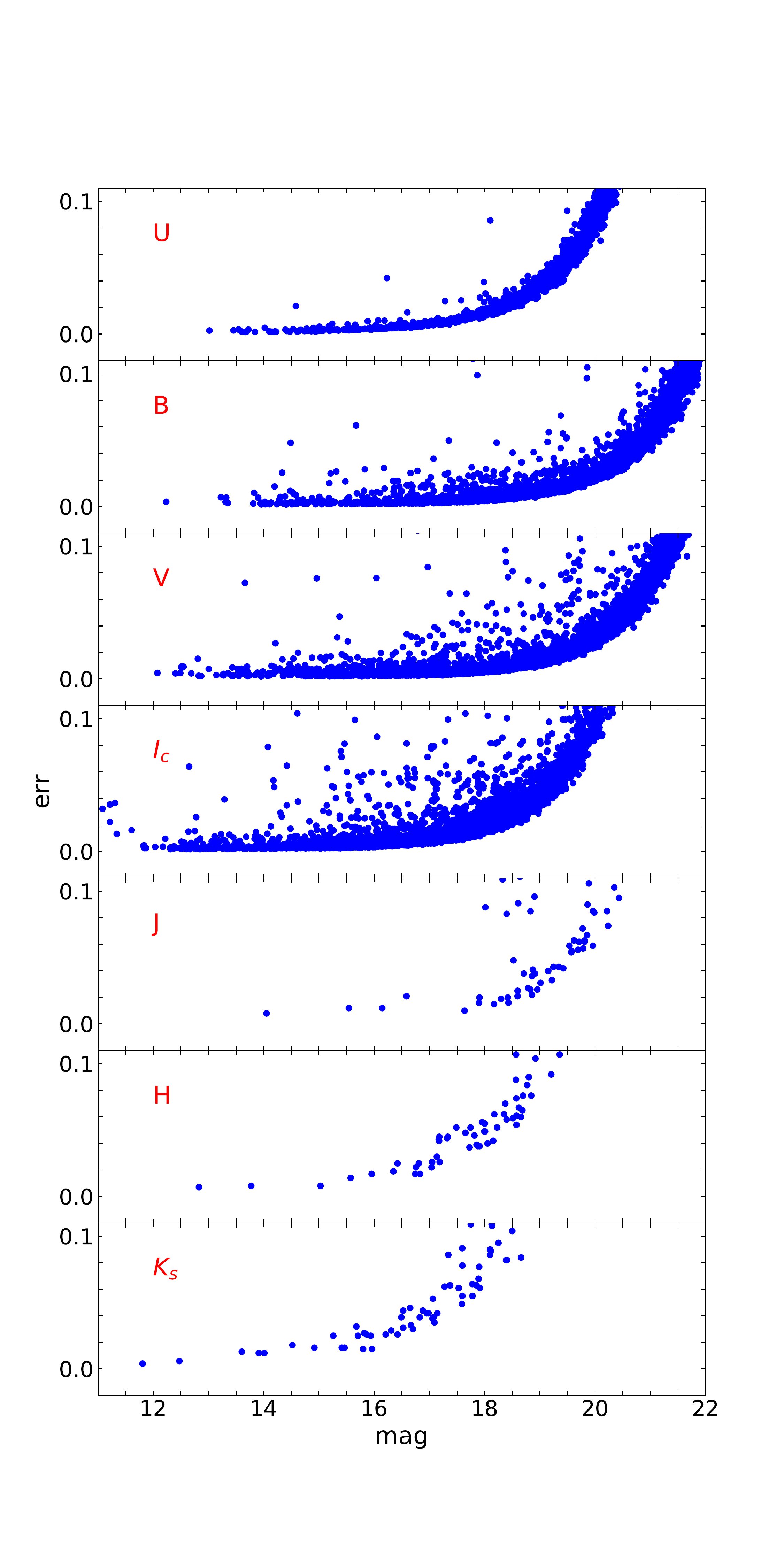}
    \caption{Photometric error versus magnitude plots in $U, B, V, I_c$ bands for 1.3m DFOT having a FOV of $18^\prime.5 \times18^\prime.5$ and in $J, H$ and $K_s$ bands for TANSPEC on 3.6m DOT having a FOV of $1^\prime \times1^\prime$.}
    \label{fig:optical_mag}
\end{figure}

Here, $U$, $B$, $V$ and $I_c$ represent the standard magnitudes whereas $u$, $b$, $v$ and $i_c$ represent the instrumental magnitudes normalized for corresponding exposure time; and $X_U$, $X_B$, $X_V$ and $X_I$ represent the airmass in corresponding bands. The left panel of Figure \ref{fig:optical_mag} represents the error versus magnitude plots for these bands.\\

The extracted magnitudes are also compared with archive `APASS'\footnote{The AAVSO Photometric All-Sky Survey, https://www.aavso.org/apass}.The difference between APASS and present standard magnitudes is negligible, i.e.,  $0.01 \pm 0.06$ mag and $0.06 \pm 0.10$ mag, for $V$ and $B$ bands, respectively.
We have reached up to detection limits of 21.66, 21.86, 20.19, 20.31 in $U$, $B$, $V$ and $I_c$ bands respectively having photometric error $\leq0.1$ (cf. Figure \ref{fig:optical_mag}).

\subsection{NIR Photometric Observation and Data Reduction}
NIR observations complement optical measurements, these can be employed to extend optical observations to the most opaque regions of molecular clouds as longer wavelength can have less extinction. NIR photometric observation of region R1 and R2 (cf. Figure \ref{fig:intro_rgb}) was done using the TIFR-ARIES Near Infrared Spectrometer (TANSPEC; \citealt{2022PASP..134h5002S}) which is mounted on the main port (Cassegrain focus) of the 3.6m Devasthal Optical Telescope (DOT) and has a FOV of $1^\prime \times1^\prime$ (with plate scale = 0.245 arcsec/pixel) which is well suited for the area of the cluster region Kronberger 55 (cf. Section \ref{sec:structure}). The observation was carried out in seven-point dithered pattern in $J$, $H$ and $K_s$ bands with 10 seconds exposure time. The total integration time for the observation was 35 mins for each band. Few dark and flat frames were also taken during the same night. Sky frames were generated by median combining the dithered frames in $J$, $H$ and $K_s$ bands individually. The basic image processing and photometry was done using the packages available within Image Reduction and Analysis Facility (IRAF). Due to crowded target field, we perfomed PSF photometry in order to obtain the instrumental magnitudes. For the calibration purpose, we have selected only isolated sources and then compared their instrumental magnitudes with 2MASS. The obtained calibration equations are as follows (see also \citealt{Sharma_2022}):

\begin{equation}
    \begin{split}
        (J-H) = (0.86 \pm 0.16) \times (j-h) - 0.09 \pm 0.21,
    \end{split}
\end{equation}
\begin{equation}
    \begin{split}
        (H-K_s) = (1.17 \pm 0.28) \times (h-k_s) + 0.76 \pm 0.12,
    \end{split}
\end{equation}
\begin{equation}
    \begin{split}
        (J-K_s) = (1.00 \pm 0.14) \times (j-k_s) + 0.51 \pm 0.20,
    \end{split}
\end{equation}
\\and\\
\begin{equation}
    \begin{split}
        (J-j) = (-0.10 \pm 0.14) \times (J-H) - 2.80 \pm 0.14
    \end{split}
\end{equation}
where, $J,H$ and $K_s$ are the standard magnitudes of the stars taken from 2MASS whereas $j,h$ and $k_s$ are the instrumental magnitudes from TANSPEC observations. Figure \ref{fig:optical_mag} represents the error versus magnitude plots for the calibrated magnitudes. We can reach 20.5 mag, 19.5 mag, and 18.5 mag in $J$, $H$, and $K_s$ bands, respectively for the stars having S/N $>$10 (error $<$ 0.1 mag).
For comparison, we have shown the images taken from the DFOT, 2MASS and TANSPEC observations for the Kronberger 55 cluster region in Figure \ref{fig:intro_rgb_comparison}. Clearly, the TANSPEC observations (stellar  profile FWHM$\sim$0.7 arcsec) is much deeper and can resolve faint stars as compared to the DFOT/2MASS observations (stellar profile FWHM$\sim$ 2 arcsec). 
 Here, it is worthwhile to note that this cluster is detectable in NIR bands but not in optical bands (cf. Figure \ref{fig:intro_rgb_comparison}), therefore, it might be a young cluster still embedded in its parental molecular cloud \citep{2003ARA&A..41...57L}.

\subsubsection{Completeness of the NIR Photometric Data}

\begin{figure}
    \centering
    \includegraphics[width=0.48\textwidth]{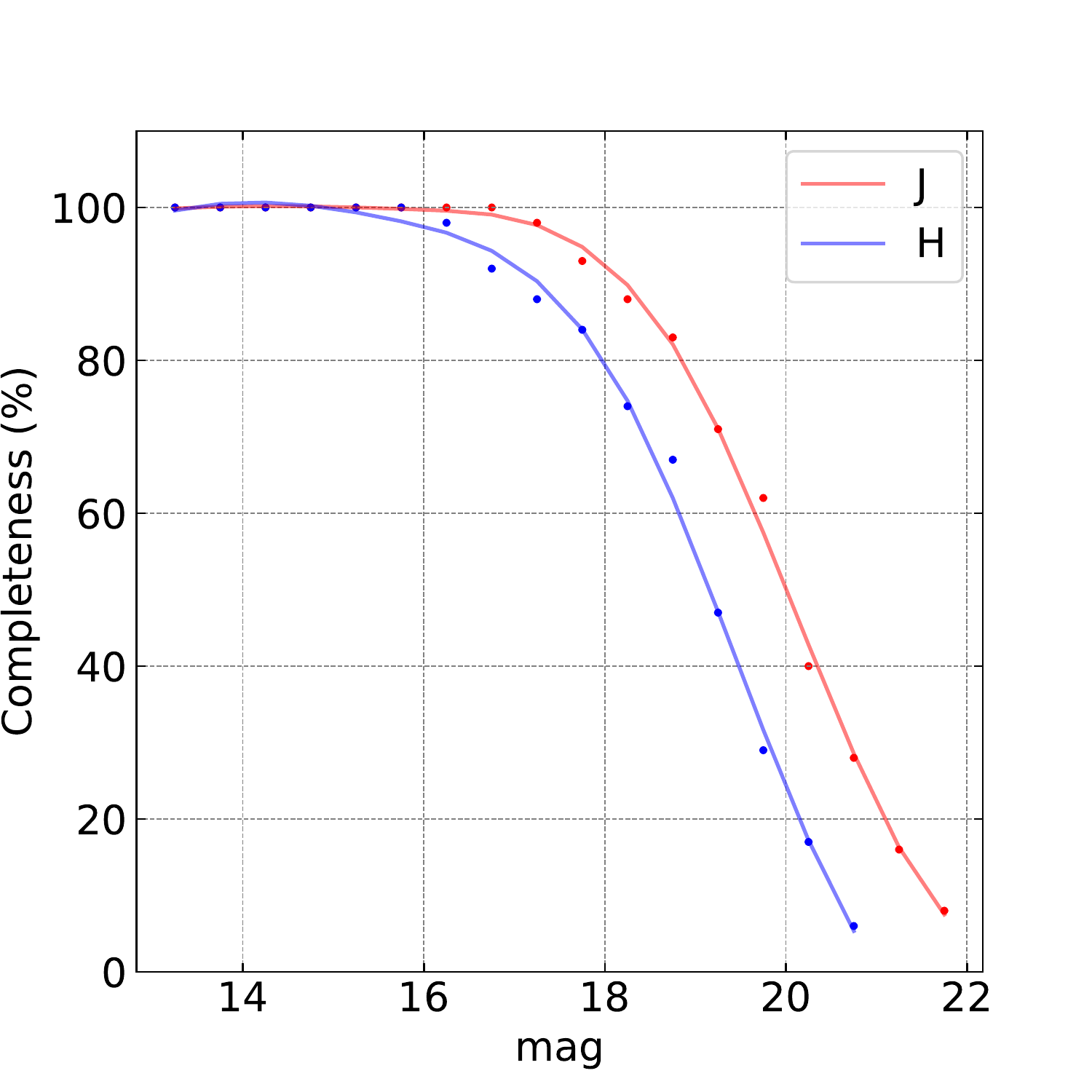}
    \caption{Completeness factor as a function of magnitude derived from the artificial star experiments (ADDSTAR).}
    \label{fig:completeness}
\end{figure}

The obtained photometric data might not be complete due to several components, e.g., crowding of stars, detection limit of the instrument, nebulosity etc. Thus, it becomes pivotal to determine the completeness of the data. In order to do so, we evaluate a factor called completeness factor (CF) through ADDSTAR routine in DAOPHOT enforced in the IRAF software package by introducing some artificial stars (for details, refer, \citealt{2008AJ....135.1934S}). The frames generated through ADDSTAR routine are then reduced with the same parameters as that of the original frames. CF is given as the ratio of total number of stars recovered to the total number of added stars in different magnitude intervals. Figure \ref{fig:completeness} show the completeness in $J$ and $H$ bands as a function of magnitude. We reach up to a detection limit of 18.9 mag and 18.0 mag in $J$ and $H$ bands, respectively, with CF $\ge 80\%$.\\ 

\subsection{Archival Data}
We have used various archival data sets (images and catalogs), such as: The NRAO VLA Sky Survey\footnote{https://www.cv.nrao.edu/nvss/postage.shtml}  (NVSS; \citealt{1998AJ....115.1693C}), \emph{Herschel}\footnote{http://archives.esac.esa.int/hsa/whsa/}\footnote{http://www.astro.cardiff.ac.uk/research/ViaLactea/} \citep{2010PASP..122..314M}, \emph{Spitzer}\footnote{https://irsa.ipac.caltech.edu/data/SPITZER/GLIMPSE/overview.html}, 2MASS\footnote{https://irsa.ipac.caltech.edu/Missions/2mass.html} \citep{2006AJ....131.1163S}, Wide-field Infrared Survey Explorer\footnote{https://irsa.ipac.caltech.edu/Missions/wise.html} (\emph{WISE}; \citealt{2010AJ....140.1868W}) and Canadian Galactic Plane Survey\footnote{https://www.cadc-ccda.hia-iha.nrc-cnrc.gc.ca/en/cgps/https://www.cadc-ccda.hia-iha.nrc-cnrc.gc.ca/en/cgps/} (CGPS; \citealt{2003AJ....125.3145T}) for the characterization and identification of YSOs and to get the impression of star formation in the region. The sources with photometric error $\leq$ 0.1 in the photometric catalogs are used for our further analysis.\\ 
\section{Result and Analysis}\label{sec:result}
\subsection{Structure of the Kronberger 55 Cluster}\label{sec:structure}
To get an insight of the structure of the Kronberger 55 open cluster, a stellar number density map was generated for the sample of stars from 2MASS covering an area of $18^\prime.5 \times18^\prime.5$. The stellar surface density maps have been generated by applying the nearest neighbour (NN) method as explained in \citet{2016AJ....151..126S} and also in \citet{2005ApJ...632..397G}. The radial distance has been varied such that it encompasses the 20th nearest star with a grid size of 6$^{''}$. In this method, at each sample position [i, j] in a uniform grid, the projected radial distance $r_N(i,j)$ to the Nth nearest star is measured. For the present study, we have taken N=20. The local stellar surface density $\rho(i,j)$ at each grid position [i, j] is computed as follows:
\begin{equation}
    \begin{split}
      \rho(i,j) = \frac{N}{\pi r_N^2(i,j)}.  
    \end{split}
\end{equation}

The Figure \ref{fig:intro_rgb} represents the color-composite image of the region generated using \emph{WISE} 22 $\mu$m (red), \emph{spitzer} 3.6 $\mu$m (green) and \emph{2MASS} K (2.17 $\mu$m) (blue) band images which has been overlaid with the stellar surface density maps (cyan contours) and the NVSS 1.4 GHz radio continuum contours (white contours). The lowest level for the isodensity contours is 4$\sigma$ above the mean stellar density (i.e., 3 stars arcmin$^{2}$) with a step size of 1.25 stars arcmin$^{2}$ whereas for the generation of radio map, it is 1.20 mJy with a step size of 0.186 mJy. From the stellar density contours, it is evident that Kronberger 55 is a small cluster of stars with circular morphology. The center of this cluster is $\alpha_{J2000}=23^h53^m10^s$, $\delta_{J2000}=+62^{0}46^{'}55^{''}$ and the radius is $0.76^{'}$.

The \emph{WISE} 22 $\mu$m emission near Kronberger 55 cluster (cf. Figure \ref{fig:intro_rgb}) is an indicative of the distribution of warm gas and dust may be due to the feedback from young massive star$/$s, whereas \emph{WISE} 12 $\mu$m hints towards poly-aromatic hydrocarbon (PAH) emission featuring at 11.3 $\mu$m, these are strong indications towards star formation in this region \citep{2020ApJ...896...29K, 2020ApJ...905...61P}. Furthermore, the MIR emission at 3.6 $\mu$m wavelength points towards the distribution of gas and dust. The diffuse radio emission (ionised regions) towards the south direction of Kronberger 55  cluster suggests the presence of massive star/s having strong UV emission.
 Therefore, it seems that Kronberger 55 is a young  cluster embedded in the nebulosity of a molecular cloud showing star formation activities.
        
        \begin{figure*}
        \centering
        \includegraphics[width=0.45\textwidth]{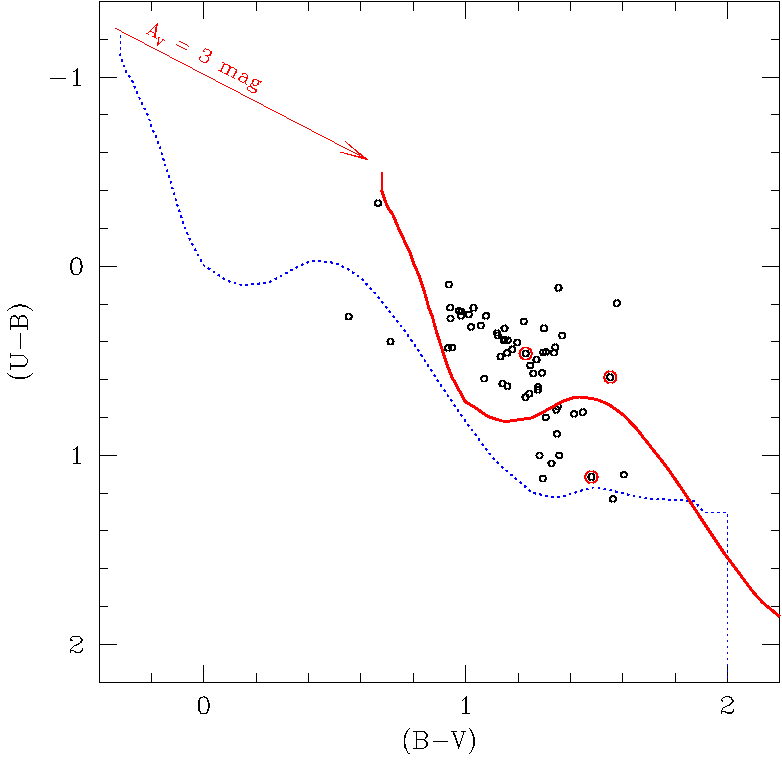}
        \includegraphics[width=0.45\textwidth]{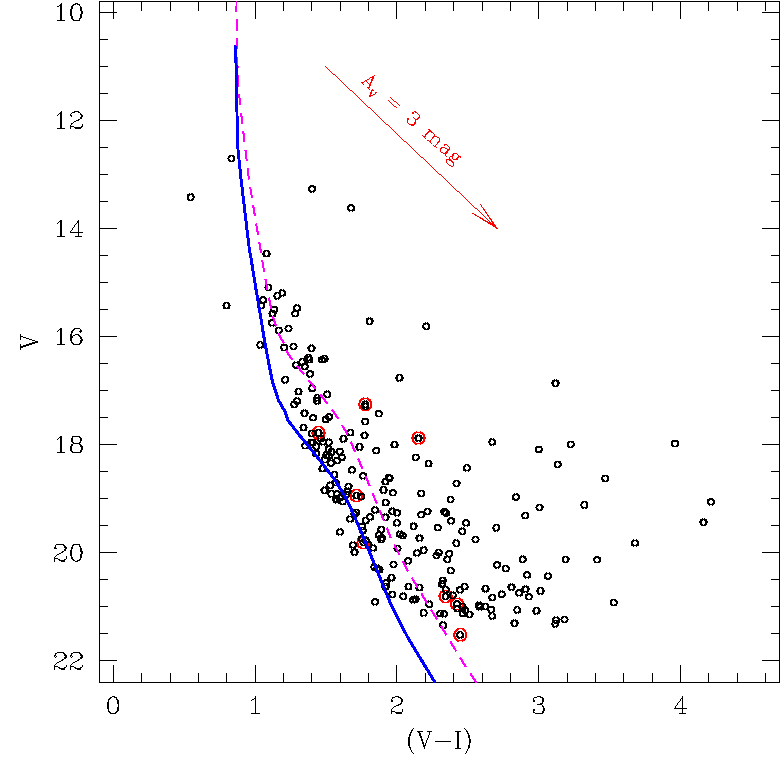}
        \caption{\label{fig:cmd}
        {\it Left panel}: $(U-B)$ vs. $(B-V)$ TCD for all the optically detected
        sources in the Kronberger region ($r_{active~region}<3^\prime$).
        Red open circles are cluster member stars identified within $r_{active~region}<3^\prime$.
        The dotted blue curve represents the intrinsic Zero Age Main
        Sequence (ZAMS) for $Z=0.02$ by \citet{Pecaut_2013}.
        The continuous red curve represents ZAMS shifted along
        the reddening vector
        for $E(B-V)$ = 1 mag.
        {\it Right panel}: $V$ vs. $(V-I_c)$ CMD for similar sources.
         The ZAMS (\citealt{Pecaut_2013}) (magenta dashed curve corrected for the distance of 2.0 kpc and blue solid curve corrected for the distance of 3.5 kpc)  corrected for reddening of $E(B-V)=1$ mag are also shown.
        }
        \end{figure*}
        
        \begin{figure*}
        \centering
        \includegraphics[width=0.75\textwidth]{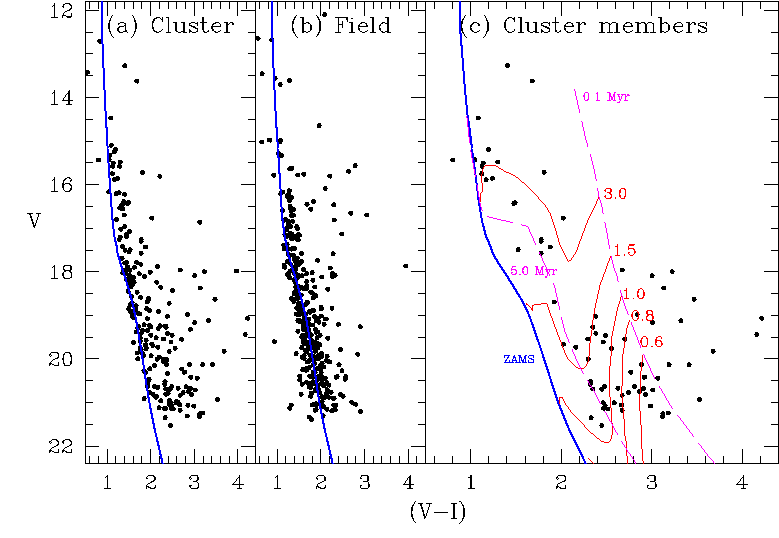}
        \caption{$V$ versus $(V-I)$ CMD for (a) stars found within Kronberger 55 $r_{active~region}<3^\prime$, (b) stars found within field or reference region, and (c) statistically cleaned $V$ versus $(V-I)$ CMD which are over-plotted with the isochrone (solid blue curve) taken from \citet{2019MNRAS.485.5666P} for 5 Myr, and the evolutionary tracks (solid red curves) of different masses taken from \citet{2000A&A...358..593S}. All the isochrones have been corrected for a distance of 3.5 kpc and reddening $E(B-V)=$1 mag.} 
        \label{fig:cmd2}
        \end{figure*}

\begin{figure}
    \centering
    \includegraphics[width=0.45\textwidth]{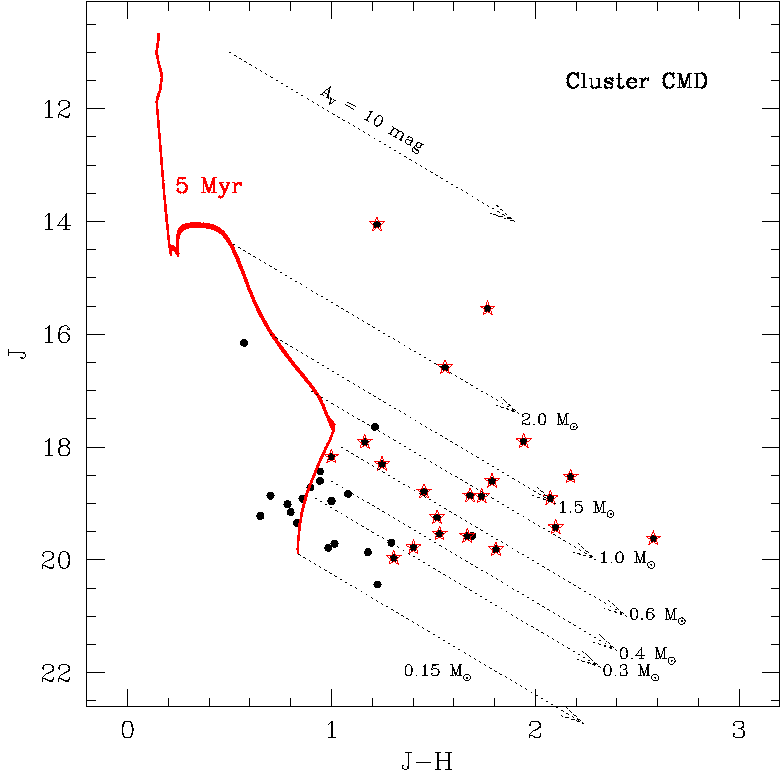}
    \caption{$J$ versus $(J-H)$ CMD generated using TANSPEC data for the stars lying within Kronberger 55 region. The red solid curve represents the isochrone for 5 Myr age taken from \citet{2019MNRAS.485.5666P}. The isochrone has been corrected for a distance of 3.5 kpc and extinction ($(E(B-V)$ = 1.0 mag). Young  pre-main sequence stars with excess IR-emission are shown with red asterisks.}
    \label{fig:cmd3}
\end{figure}

\subsection{Extinction, distance and age of the cluster}

Since we were not able to find the optical counterparts of the sources in Kronberger 55 cluster ($1^\prime \times1^\prime$ particularly), so we have selected a bigger `active' region showing star formation activities based on the distribution of gas and dust surrounding the Kronberger 55 clusters ($r_{active~region} < 3^\prime$, refer Figure \ref{fig:intro_rgb})  for the extraction of various parameters using optical data.

We have used $(U-B)$ versus $(B-V)$ optical TCD to estimate the extinction towards the cluster Kronberger 55 (cf. left panel of Figure \ref{fig:cmd}). We have shown the distribution of the stars lying within active region with black dots along with the zero-age main sequence (ZAMS, the blue dotted curve) which we have taken from \citet{Pecaut_2013}. We have also shown the stars located within the Kronberger 55 cluster boundary by red circles. Since there is a large spread in the distribution of the stars along the reddening vector, it indicates differential reddening in the region. The ZAMS has been shifted along the reddening vector by a slope of $E(U - B)/E(B - V)$ = 0.72 (which corresponds to $R_V=3.1$) to attain the distribution of star representing minimum reddening along the cluster direction. For reddening analysis, we have chosen only those stars whose spectral type is A or earlier. There are several factors which have led to this choice, such as, distribution of binary stars, error in photometry, metallicity, pre-main sequence (PMS) stars, rotation (see \citealt{1974ASSL...41.....G, 1994ApJS...90...31P}). In such a way, the minimum reddening value ($E(B-V)_{min}$) comes out to be 1.0 mag which has been shown by red curve. The approximate error in this estimation is 0.05 mag. The procedure to estimate error has been outlined in \citet{1994ApJS...90...31P}.

\begin{figure*}
    \centering
    \includegraphics[width=0.48\textwidth]{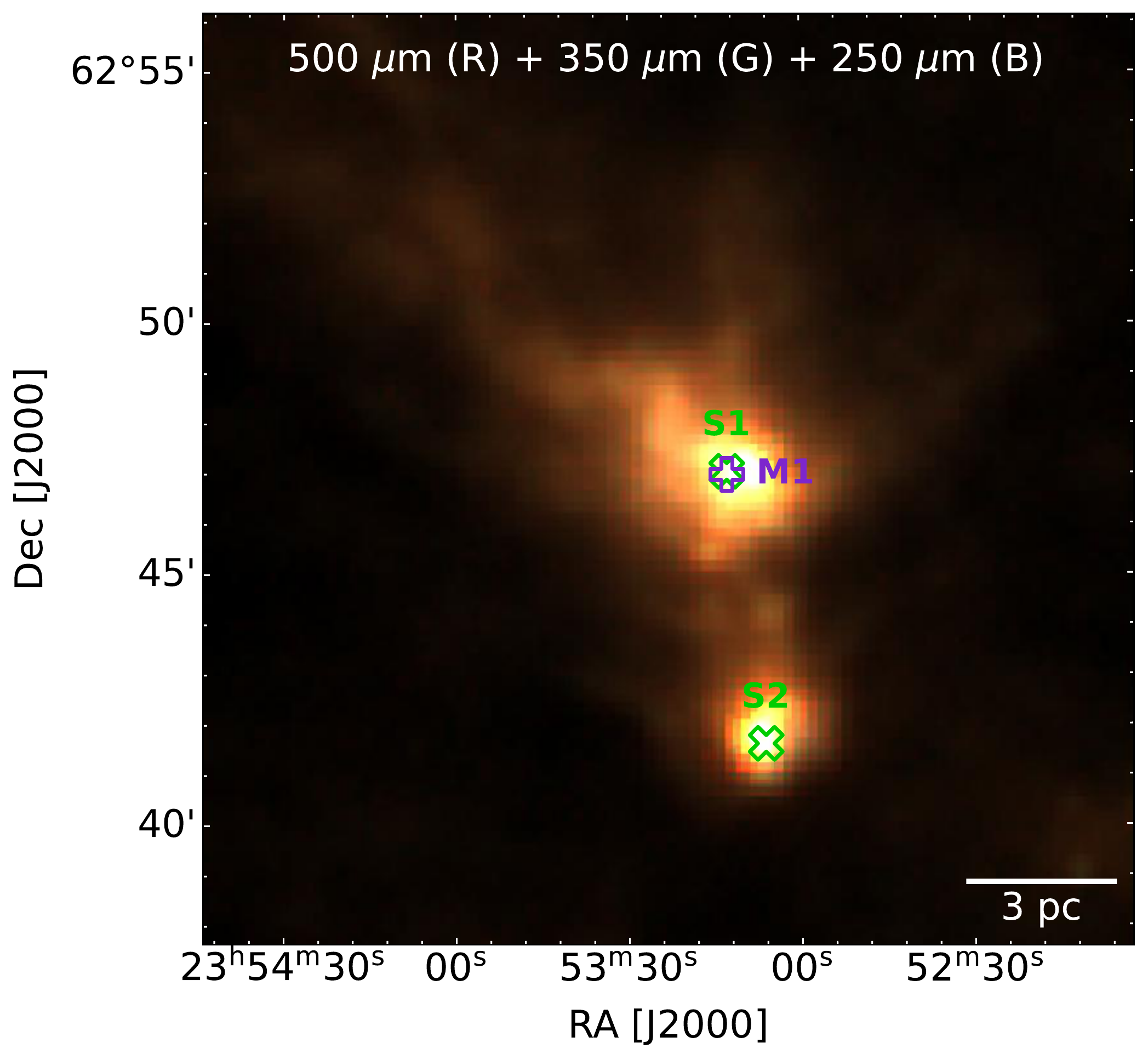}
    \includegraphics[width=0.49\textwidth]{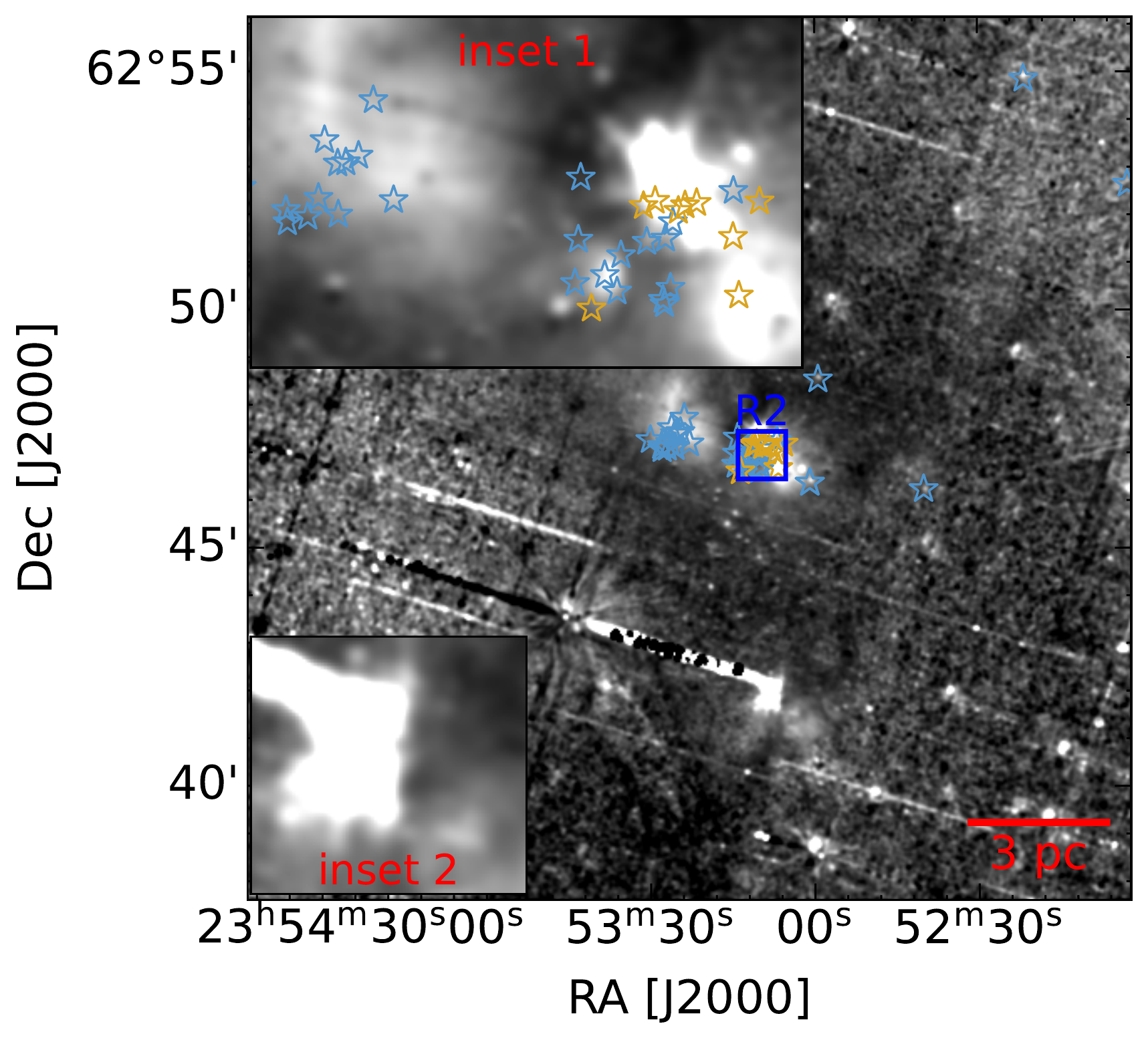}
    \includegraphics[width=0.48\textwidth]{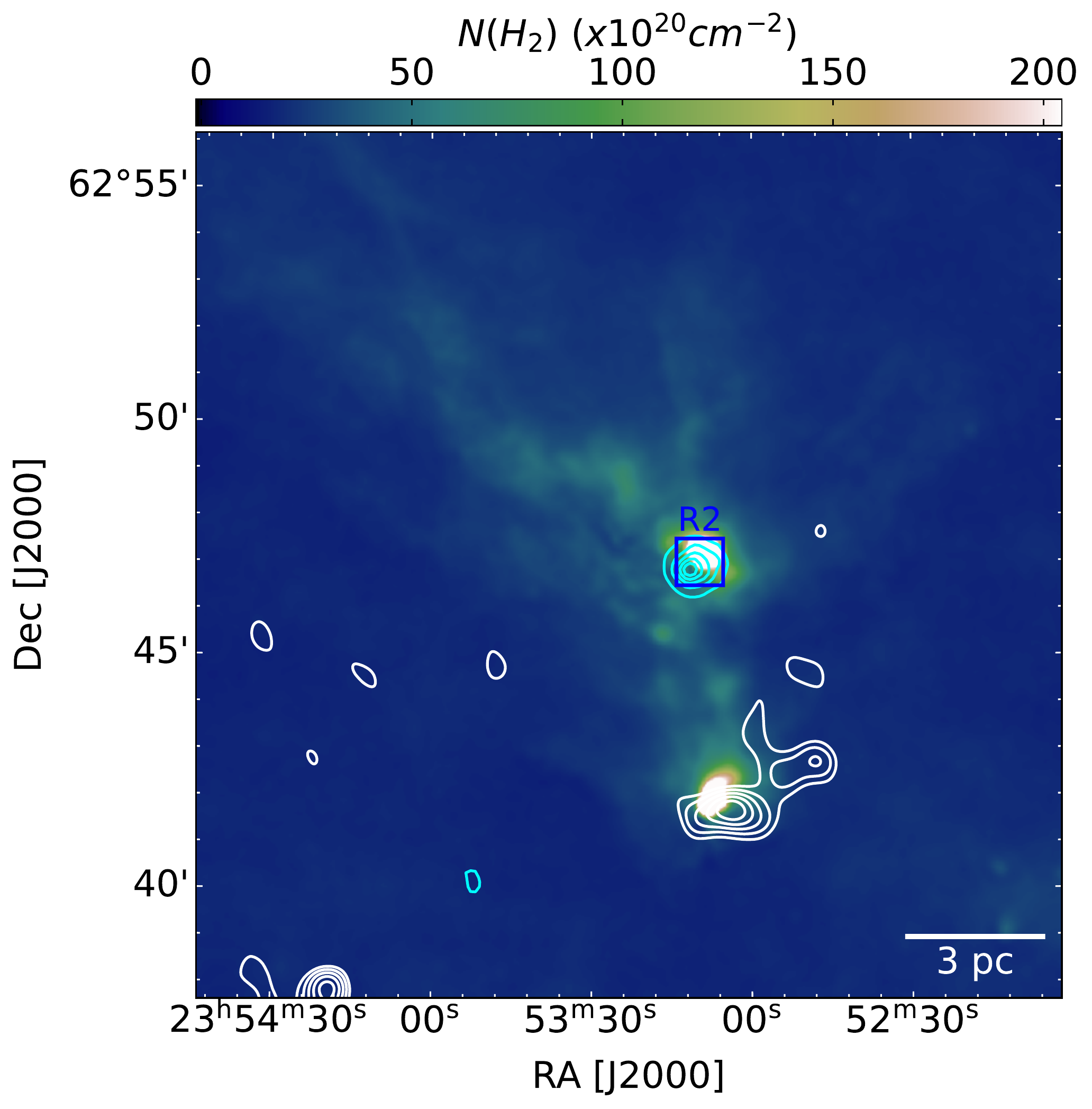}
    \includegraphics[width=0.48\textwidth]{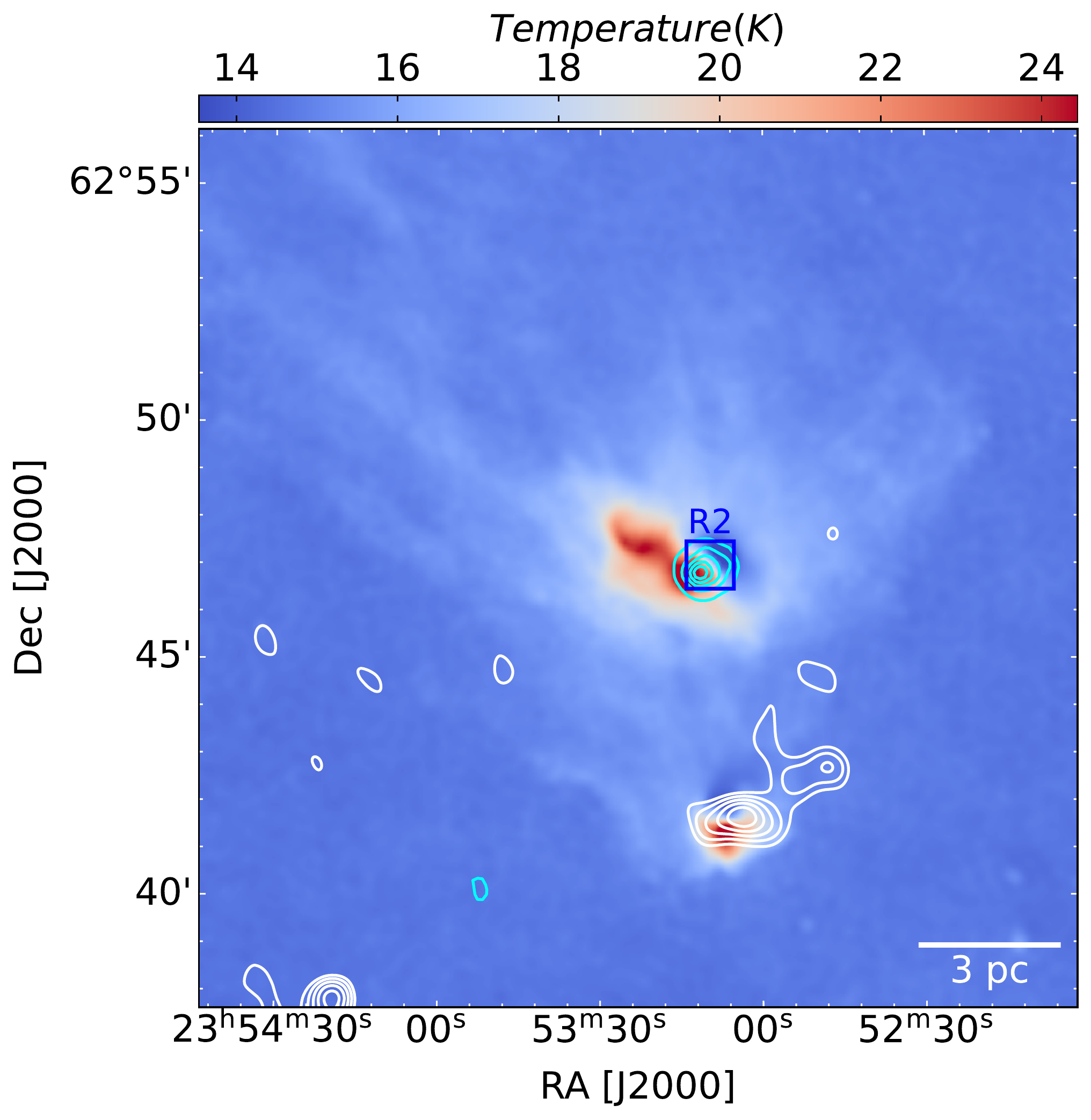}
    \caption{The upper left panel: Color-composite image (Red: \emph{Herschel} 500 $\mu$m; Green: \emph{Herschel} 350 $\mu$m and \emph{Herschel} 250 $\mu$m) of the region overlaid with the locations of IRAS sources S1 and S2 (shown by green `$\times$' along with an H $\textsc{ii}$ region M1 (shown by magenta $`+$' which overlaps with S1. Upper right panel: \emph{Spitzer} ratio map of $4.5 \mu m / 3.6 \mu m$ emission, smoothened using Gaussian function with a radius of two pixel. The location of identified Class $\textsc{I}$ and Class $\textsc{II}$ YSOs are also marked by yellow and blue asterisks respectively. The inset 1 represents the zoomed-in view of the ratio map where Kronberger 55 is located,  whereas the inset 2 represents the zoomed-in view of the ratio map where radio emission takes place. Lower left panel: \emph{Herschel} column density map. Lower right panel: \emph{Herschel} temperature map ovrlaid with the isodensity contours (cyan) generated from the 2MASS catalog and NVSS 1.4 GHz radio continuum contours (white). The blue square represents R2.}
    \label{fig:rgb_images}
\end{figure*}

Many previous studies have reported different values of distance and  age for this cluster (as described in Section \ref{sec:intro}). In this study, we have derived these values by using the classical method of isochrone fitting by using the optical CMD \citep{Sharma_2006, 2014A&A...563A.117F, 2015A&A...576A...6P, 2019A&A...623A.108B, 2020ApJ...891...81P, 2020ApJ...896...29K}. The right panel of the Figure \ref{fig:cmd} represents the $V$ versus $(V-I)$ CMD for the stars lying within the active region. The stars located within the cluster boundary are shown with red circles. 
The dashed magnta curve is ZAMS \citep{Pecaut_2013} corrected for a distance of 2 kpc and $E(B-V)$ = 1.0 mag \citep{2016A&A...585A.101K}. Clearly, this distance value seems to be under-estimated as usually ZAMS is fitted to the lower envelop of the MS. The ZAMS seems to be better fitted to the lower envelope of the MS at a distance of 3.5 kpc (blue solid curve).  This visual fitting of ZAMS to MS is done where there is a bend in the MS, it is based upon various factors, e.g., evolutionary effects, distribution of binary stars, rotation etc (refer \citealt{1974ASSL...41.....G, 1994ApJS...90...31P} for further details). Earlier \citet{2013yCat..22080011L} and \citet{2021yCat..22540003M} have estimated the distance of G116.327+00.639-045.90 molecular cloud and an H $\textsc{ii}$ region G116.3282+00.6627, which are located (marked as M1 in the upper left panel of Figure \ref{fig:rgb_images}) exactly at the same position as that of Kronberger 55, at a distance of 3.5 kpc, which is best matched with our estimated distance. 

We have also tried to estimate the age of the cluster by isochrone fitting to the CMD of cluster stars \citep{2007MNRAS.380.1141S}.
Since this cluster seems to be of young age (see Section \ref{sec:structure}), probability of finding young PMS low-mass stars are high in this cluster. The age can thus be constrained based on the pre-main sequence turn-on point or fraction of stars with disc to  those without discs stars in the cluster. Since, for the latter, there is no well calibrated relation in the literature we will concentrate on the PMS turn-on point in the CMD. To get this point we need a clean sample of stars belonging to cluster without any contamination from the foreground/background sources. Therefore, We have statistically subtracted the field CMD (having equal area as of cluster) to the cluster CMD, to get the real cluster CMD (for details please refer \citealt{2007MNRAS.380.1141S}, \citeyear{10.1093/pasj/64.5.107}, \citeyear{2017MNRAS.467.2943S}; \citealt{2013MNRAS.432.3445J}).
In the left-panel of the Figure \ref{fig:cmd2}, we have shown $V$ versus $(V-I_c)$ CMD for the stars located within the active region, in the middle-panel, we have shown the CMD for reference field region ($7^{\prime}.2<r<7^{\prime}.8, r$ is the distance from the center of the cluster). In the right-panel, we have plotted the statistically cleaned $V$ versus $(V-I_c)$ CMD for the cluster region which clearly indicates the presence of PMS stars.
The dashed magenta curves in the figure are PMS for 0.1 and 5 Myrs and the red curves are the PMS for stars having different masses \citep{2000A&A...358..593S}. The isochrones are corrected for a distance of 3.5 kpc and reddening of $E(B-V)$ = 1 mag. From this distribution, it seems that the most of the stars in the active region are low mass PMS stars having ages less than 5 Myrs.

To further confirm the age of stars in this region, we have plotted in Figure \ref{fig:cmd3} the $J$ versus $(J-H)$ CMD of the stars located in the Kronberger 55 cluster region which are detected in our deep NIR photometry on the TANSPEC data taken with 3.6m DOT. We have also shown the location of young pre-main sequence stars showing excess IR emission (cf. \ref{appendixA}).
Most of the disc bearing stars have ages 0.5 - 3 Myr \citep{Evans_2009}.
We can clearly see the distribution of low-mass young PMS stars age embedded in the nebulosity of this regions. Thus, confirms our age estimation (age $\lesssim$ 5Myr) of Kronberger 55 cluster.

\subsection{Physical environment around Kronberger 55 cluster}\label{sec:physical_environment}

The high-resolution IR, submm and radio observations help us to better understand the distribution of young stars, gas, dust and ionized gas, etc. \citep{2010A&A...523A...6D, 2017ApJ...834...22D}.

The upper left panel of Figure \ref{fig:rgb_images} represents the color-composite image of the region generated using the \emph{Herschel} 500 $\mu$m (red), 350 $\mu$m (green), and 250 $\mu$m (blue) band images. These MIR images show the presence of extended structure of dust and gas emission in the region with two peaks, one at the location of cluster Kronberger 55 and another at 5$'$.35 southwards to it. Both of these peaks have also cold IRAS point sources associated with them, i.e, IRAS 23507+6230 and IRAS 23506+6224 \citep{2016yCat..17840111L} termed as S1 and S2, respectively (shown by green $'\times'$ symbol). An H $\textsc{ii}$ region `G116.3282+00.6627'  overlapping with S1 is also reported by \citet{2013yCat..22080011L} (termed as M1  and shown by magenta $`+'$ symbol).

The upper right panel of Figure \ref{fig:rgb_images} represents the \emph{Spitzer} ratio map of $4.5 \mu m / 3.6 \mu m$ emission which is smoothened using Gaussian function with a radius of two pixel. Since, 3.6 $\mu$m and 4.5 $\mu$m bands have same point spread function, so they can be divided directly (refer \citealt{2017ApJ...834...22D} for further details). This map imparts some bright and dark regions. The 4.5 $\mu$m (indicated by brighter regions in the map) holds a prominent Br$-\alpha$ emission at 4.05 $\mu$m and a molecular hydrogen line at 4.693 $\mu$m, whereas 3.6 $\mu$m (indicated by darker regions in the map) holds the PAH emission at 3.3 $\mu$m, which indicates towards the presence of photo-dissociation region (PDR), which might have been produced due to the presence of massive star$/$s.
The inset 1 of this panel represents the zoomed-in view of the ratio map where Kronberger 55 is located, whereas the inset 2 represents the zoomed-in view of the ratio map where the radio emission takes place.
We can see bright patches near the Kronberger 55 cluster and at southern radio peak, which suggests the affect of outflow activities from YSOs in the region. 

The lower left and right panels of Figure \ref{fig:rgb_images} represents the \emph{Herschel}  column density ($N_{H_2}$) and temperature map respectively (resolution $\sim$ 12$^{''}$) procured for EU-funded ViaLactea project \citep{2010PASP..122..314M} adopting the Bayesian PPMAP technique on the \emph{Herschel} data \citep{2010A&A...518L.100M} at 70, 160, 250, 350, and 500 $\mu$m wavelengths \citep{2015MNRAS.454.4282M, 2017MNRAS.471.2730M}. Both these maps have been overlaid with isodensity and radio contour (cf. Section \ref{sec:structure}) along with the region R2. The column density map shows  filamentary structures around Kronberger 55 with two peak at the location of IRAS sources. Various authors have studied the dark/molecular clouds and found that the star formation takes place in  the regions of high extinction/column density \citep{Lada_2010, 2006A&A...454..781L, 2002AJ....123.2559C, 1999A&A...345..965C, 1999ApJ...512..250L}.
Near peaks in the column density map, the temperature map also shows dust emission warmer (i.e., $T_d \sim 21-24$K) than the surroundings. The outflow activities as seen by bright patches in the \emph{Spitzer} ratio map, are coincident with the warm dust emission, thus confirms the effect of young stars in the region.

\section{Discussion}
\label{disc}

\begin{figure}
    \centering
    \includegraphics[width=0.48\textwidth]{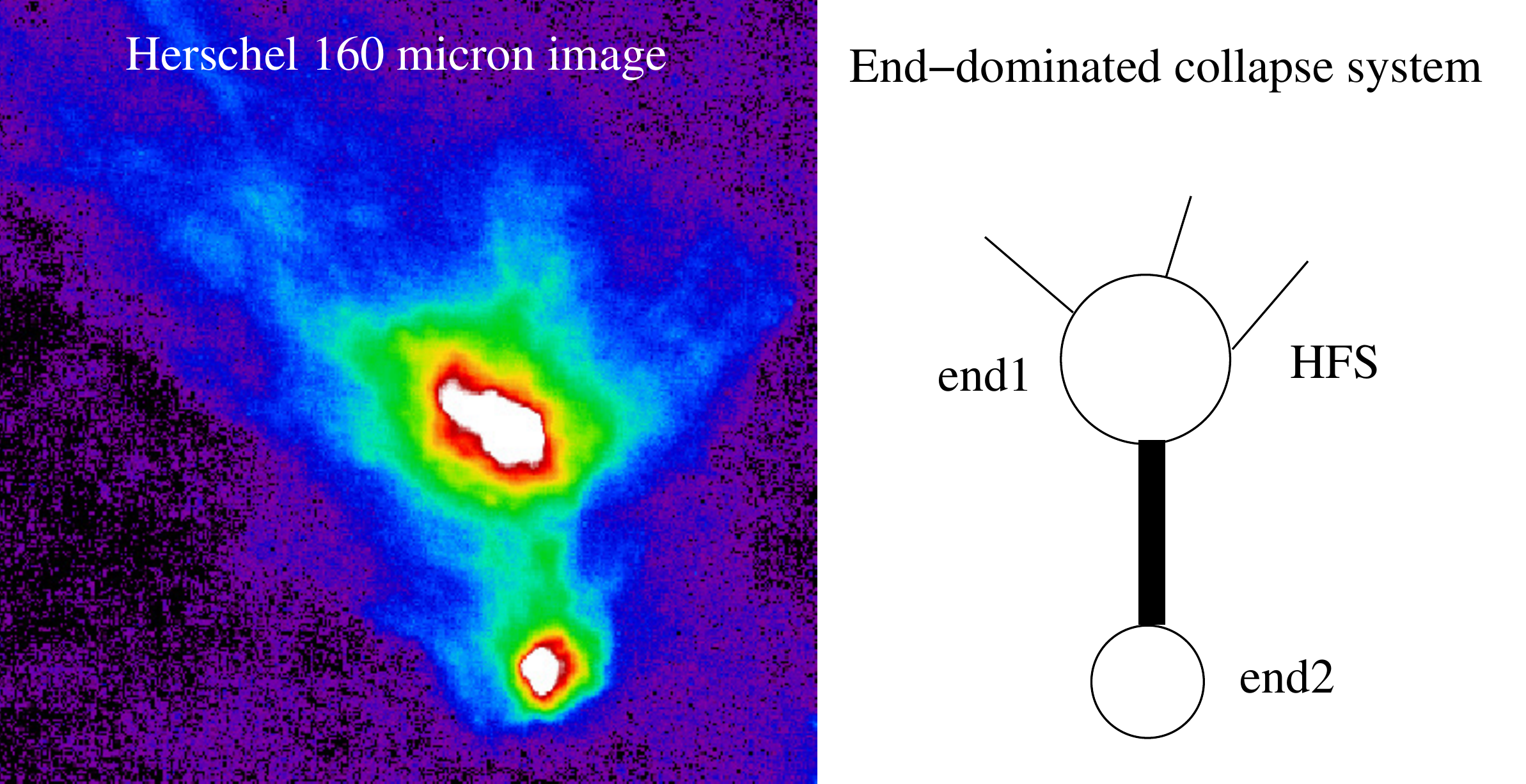}
    \caption{Left panel: The \emph{Herschel} 160 $\mu$m image of the region showing two dominant peaks connected through a filamantory structure. Right panel: Cartoon diagram depicting EDC scenario in a hub-filamentory system.}
    \label{fig:cartoon}
\end{figure}

\begin{figure}[!h]
    \centering
    \includegraphics[width=0.45\textwidth]{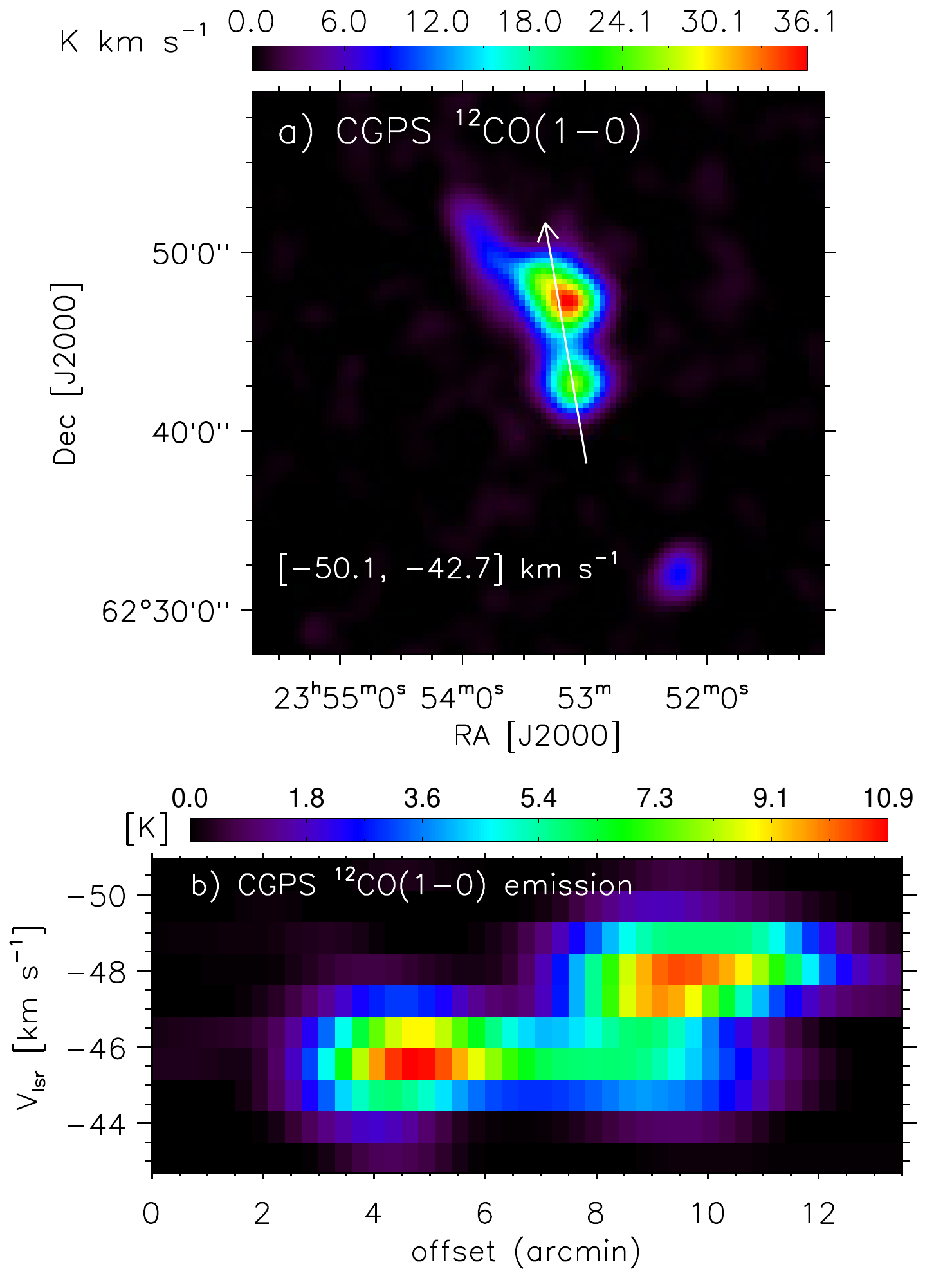}
    \caption{(a) The CGPS $^{12}$CO$(1-0)$ molecular emission map integrated over a velocity range of [-50.1, -42.7] km s$^{-1}$. (b) $pv$ diagram along the white arrow marked in panel (a). The zero-point in this diagram corresponds to the southern end of the arrow.}
    \label{fig:co_data}
\end{figure}

The Kronberger 55 cluster is found to be located in a region showing signatures of recent/active star formation activities.  It is estimated to be located at a distance of 3.5 kpc, similar to that reported for molecular clouds in the region \citep{2021yCat..22540003M}.
This confirms association of this cluster with the  distribution of molecular clouds. The cluster is also found to be of very young age, i.e., $\lesssim$ 5 Myrs and has young disc bearing PMS stars with excess IR-emission.
Matter from the disc of these stars is accreted on to the stellar surface and at certain instances, they show enhanced accretion rates or jets/outflow activities \citep{2011MNRAS.415.3380O}.
As we can see bright patches in the $Spitzer$ ratio map, it suggest excess 4.5 $\mu m $ emission or Br$\alpha$/H$_2$ emission due to the shocked region produced by the outflow activities of young stars.

There is also distribution of cold/warm gas and dust in the region as seen in the \emph{WISE} MIR images. The \emph{Herschel} maps show similar distribution having extended emission with two strong peaks in the region. The cluster Kronberger 55 lies on one of these peaks and the other peak shows diffuse radio emission. Both these peaks also show the bright patches in the \emph{Spitzer} ratio  map, suggesting the effects of outflow activities. The temperature maps is also correlated with these bright patches, thus confirming the affect of young stars in the region. The radio emission in the southern peak hints towards the formation of massive star/s. The Kronberger 55 cluster lack radio emission suggesting the formation of low mass stars in there. This is further confirmed by the distribution of low-mass young stars in the cluster region through our deep NIR observations.

This region seems to have two peaks in the molecular structure with surrounding extended regions. These extended regions, to some extent, looks like filamentary structures (a region of low aspect ratio and high column density) identified in various other regions in \emph{Herschel} maps (see \citealt{2022arXiv221004658D}).
Usually, the filametary structures are associated with a hub system (a region of very high column density with spherical geometry \citealt{Myers_2009}). Population of young stars (both massive and low mass stars) have been reported in these hubs in our Galaxy (\citealt{Dewangan_2020}; \citealt{refId0}; \citealt{Wang_2020}).  We have also found similar distribution in our studied region but with two separate hubs showing star formation activities and are connected through a filamentary structure. This kind of distribution is a HFS 
with EDC scenario. 
 In Figure \ref{fig:cartoon}, we have shown the \emph{Herschel} 160 $\mu$m image  along with a cartoon diagram depicting EDC scenario in our studied region. Clearly, we can see two peak showing star formation activities which connected through a filamentory structure.

To further explore the existence of the observed configuration as seen in the sub-millimeter maps, we examined the CGPS $^{12}$CO$(1-0)$ line data. In this relation, we produced an integrated intensity map (or moment-0 map) of the $^{12}$CO emission, which is integrated over a velocity range of [-50.1, -42.7] km/s (cf. \citealt{2021MNRAS.506.6081D}). This velocity range is obtained from the study of the velocity profile, which is not shown in this paper. 
The moment-0 map shows the distribution of molecular emission, revealing a connection of two molecular condensations (see Figure \ref{fig:co_data}a). The physical association of the molecular condensations is also seen in the position-velocity \emph{(pv)} map (Figure \ref{fig:co_data}b), which is produced along an axis passing through both the condensations (see an arrow in Figure \ref{fig:co_data}a). Note that such molecular configuration is similar as seen in the \emph{Herschel} maps. However, due to the coarse beam size of the CGPS $^{12}$CO$(1-0)$ line data, we cannot resolve the HFS toward the northern condensation as reported in the \emph{Herschel} maps. Hence, the resolution of the CGPS line data is not enough to obtain more insights into the velocity structures of the condensations and their connecting filament. Hence, this work is a scope of our future study.

\section{Conclusion}\label{sec:conclusion}

We performed a detailed analysis of the Kronberger 55 open cluster and its surrounding region using optical observations taken with 1.3m DFOT having a FOV of $18^\prime.5 \times18^\prime.5$ and deep NIR observations taken with TANSPEC having a FOV of $1^\prime \times1^\prime$ mounted on 3.6m DOT along with multiwavelength archival data sets. We have studied the structure, environment of this cluster and derived the fundamental parameters. The major results of this study are as follows:

\begin{enumerate}
    \item The structural parameters of this cluster have been derived by using isodensity contours and found that the Kronberger 55 cluster shows circular morphology. The core radius of this cluster is found to be 0.75 arcsec ($\sim$ 0.76 pc).
    \item The $J$ versus $(J-H)$ CMD plotted using deep NIR data of TANSPEC reveals the distance, extinction and age of the cluster Kronberger 55 as $\sim$3.5 kpc, $\sim$1.0 mag and $\sim$5 Myr respectively.
    \item We have identified a total of 22 YSOs using TANSPEC on the basis of excess IR emission within Kronberger 55 cluster, out of which 9 are Class $\textsc{I}$ and 13 are Class $\textsc{II}$ which cause outflow activities around Kronberger 55 cluster.
    \item We have analyzed the physical environment within $18^\prime.5 \times18^\prime.5$ FOV around Kronberger 55 using MIR to FIR images along with NVSS radio continuum. We have found that there is emission of extended dust and gas around Kronberger 55 cluster and in southwards location to this cluster along with PAH emission. The southwards location also depicts NVSS 1.4 GHz radio emission. 
    \item The \emph{Herschel} temperature map traces shell-like structure in a temperature range of $\sim 21-24$K around Kronberger 55 and the same southwards location which shows extended dust and gas emission. High column density is also seen in these two regions in the \emph{Herschel} column density map.
    \item In the \emph{Herschel} sub-millimeter images, this region seems to have clumps with a surrounding filamentary structures. These clumps are undergoing active star formation.
    \item bf Using the CGPS $^{12}$CO$(1-0)$ line data, the molecular map at [-50.1, -42.7] km/s supports the existence of two molecular condensations and their connecting filament as seen in the \emph{Herschel} maps. But, the resolution of the CGPS line data is not sufficient to resolve the HFS toward the northern molecular condensation.
    \item Our study suggests that this region might be a hub-filamentary system  where the EDC scenario plays role in the formation of young stars.
\end{enumerate}

\section*{Acknowledgements}
We thank the anonymous reviewer for valuable comments which greatly improved the scientific content of the 
paper. We thank the staff of the 1.3m DFOT and the 3.6m DOT at Devasthal, Nainital, India. This publication makes use of data from the Two Micron All Sky Survey, which is a joint project of the University of Massachusetts and the Infrared Processing and Analysis Center/California Institute of Technology, funded by the National Aeronautics and Space Administration and the National Science Foundation. This work is based on observations made with the Spitzer Space Telescope, which is operated by the Jet Propulsion Laboratory, California Institute of Technology under a contract with the National Aeronautics and Space Administration. This publication makes use of data products from the Wide-field Infrared Survey Explorer, which is a joint project of the University of California, Los Angeles, and the Jet Propulsion Laboratory/California Institute of Technology, funded by the National Aeronautics and Space Administration.
We acknowledge the financial support of DST-INSPIRE (No.$\colon$ DST/INSPIRE Fellowship/2019/IF190550).
\vspace{-1em}

\appendix

\section{Identification of YSOs in the Region}\label{appendixA}

\begin{figure*}[!ht]
    \centering
    \includegraphics[width=0.3\textwidth]{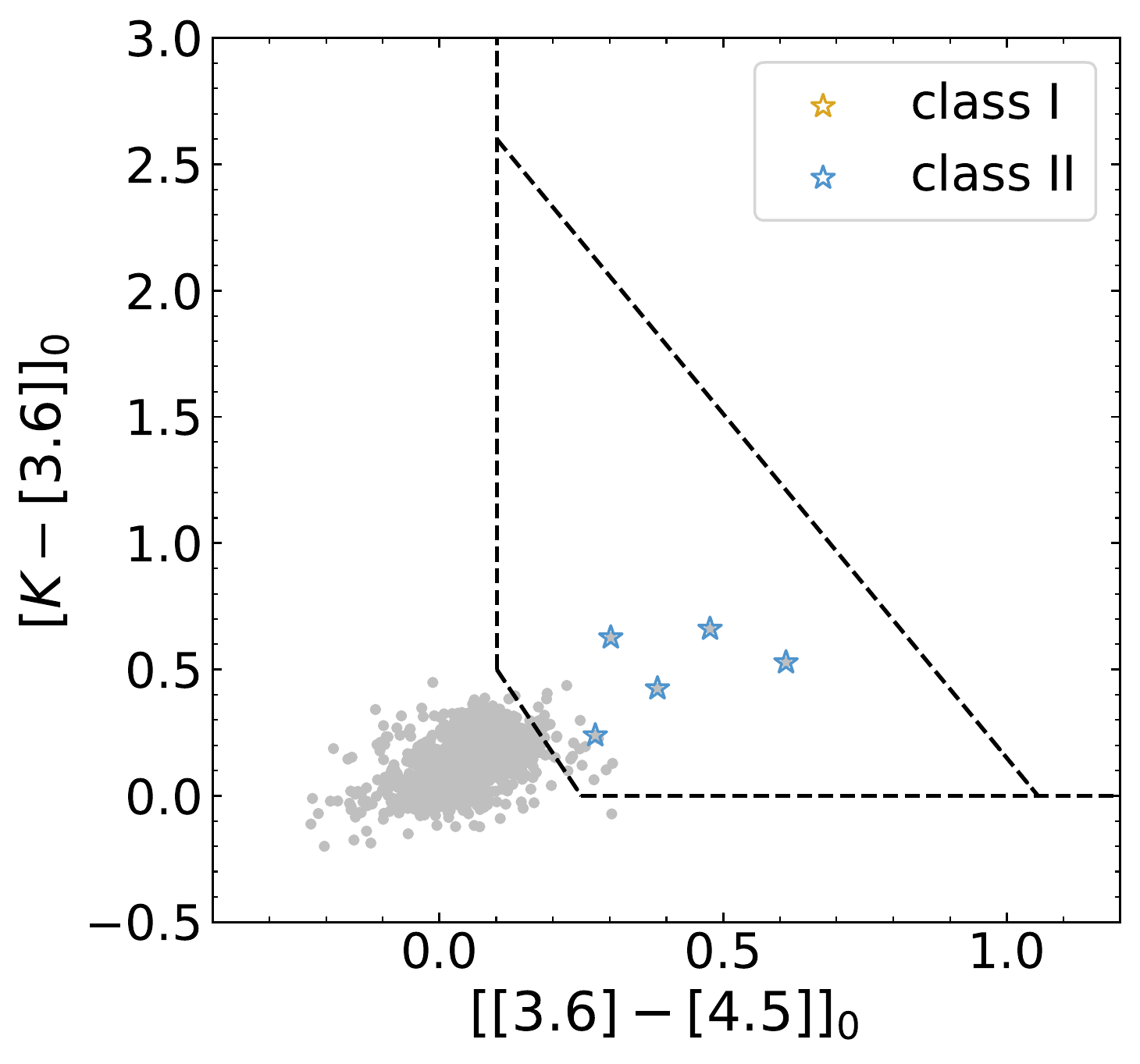}
    \includegraphics[width=0.285\textwidth]{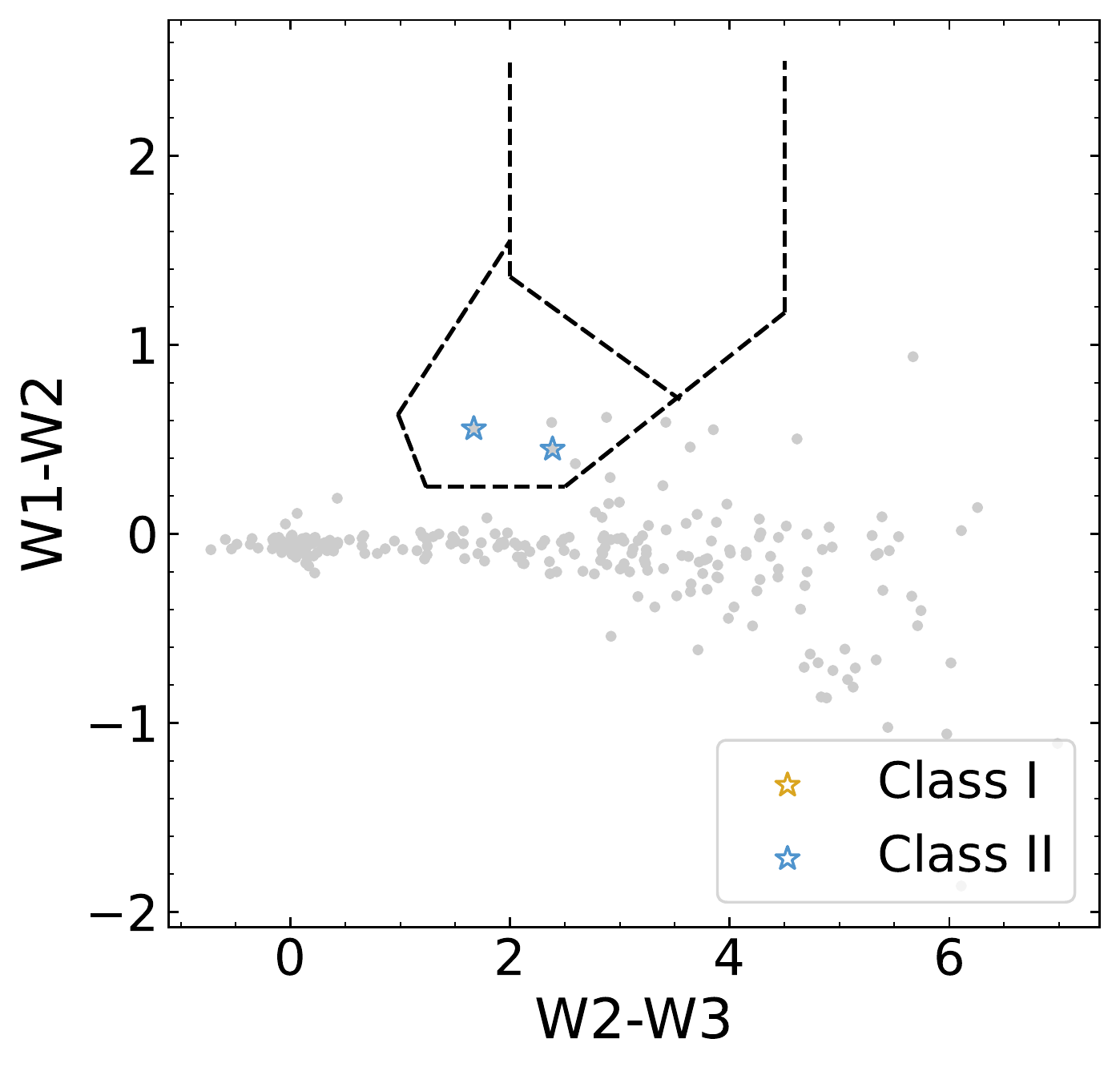}
    \includegraphics[width=0.3\textwidth]{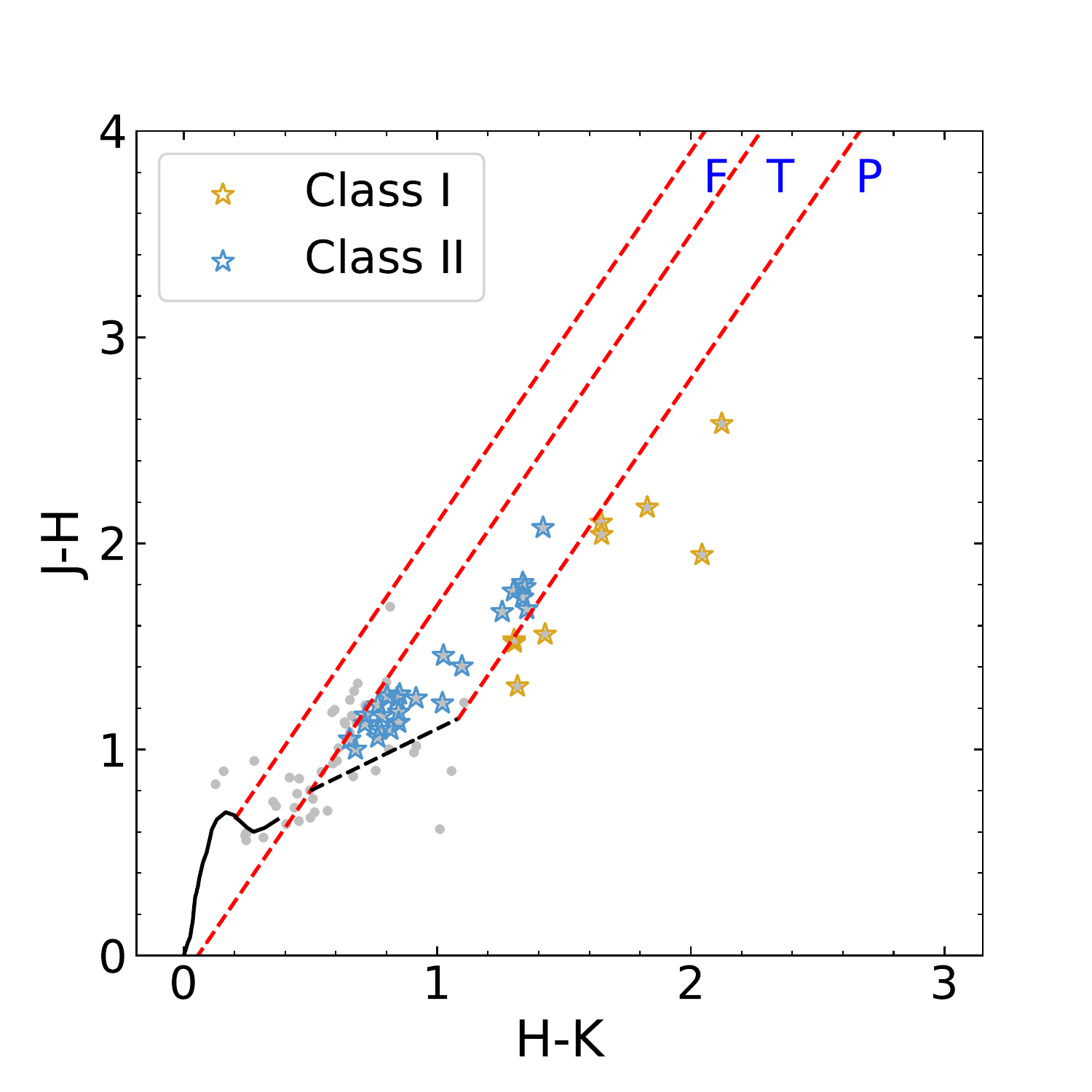}
    \caption{TCD for the YSOs identified within $18^\prime.5 \times18^\prime.5$ region. Left panel: [K - [3.6]]$_0$ vs. [[3.6] - [4.5]]$_0$ TCD, classification of YSOs is based on the scheme given by \citep{2009ApJS..184...18G}. Middle Panel: $[W1-W2]$ vs. $[W2-W3]$ TCD, classification of YSOs is based on the scheme given by \citep{2014ApJ...791..131K}. Right panel: $(J-H)$ vs. $(H-K)$ CMD for $1^\prime \times1^\prime$ Kronberger 55 cluster region, classification is based on the scheme given by \citep{2004ApJ...608..797O}. The continuous solid black and thick dashed curve represent the sequence for field dwarfs and giants respectively (taken from \citealt{Bessell_1988}) whereas the dotted line represents the locus of Classical T Tauri stars (CTTS; \citealt{1997AJ....114..288M}). Parallel dashed lines (red) represent the reddening lines \citep{1985ApJ...288..618R} drawn from the tip (spectral type M4) of the giant branch (left red line), from the base (spectral type A0) of the MS branch (middle red line) and from the tip of intrinsic CTTS line (right red line). Yellow and blue asterisks represent Class $\textsc{I}$ and Class $\textsc{II}$ YSOs respectively.}
    \label{fig:yso_classification}
\end{figure*}

Stars which are still in the PMS phase, exhibit excess IR emission due their circumstellar disk. Thus, YSOs are generally classified on the basis of their IR excess emission. We have used the \emph{GLIMPSE360} catalog of Spitzer Science Centre (SSC) by applying the updated classification scheme of \citet{2009ApJS..184...18G} and the MIR data of \emph{ALLWISE} catalog of \emph{WISE} by applying the scheme given in \citet{2014ApJ...791..131K} for the identification of YSOs. The [K - [3.6]]$_0$ vesrsus [[3.6] - [4.5]]$_0$ TCD (see left panel of Figure \ref{fig:yso_classification}) gives 5 Class $\textsc{II}$ YSOs in our region. Whereas, the ([3.4] - [4.6]) versus ([4.6] - [12]) TCD (see right panel of Figure \ref{fig:yso_classification}) gives only 3 Class $\textsc{II}$ YSOs. The YSOs having their counterparts in \emph{Spitzer} catalog, have been combined. In such way, we have obtained 6 Class $\textsc{II}$ YSOs in total.
 
We have classified the YSOs in region R1 and R2 using TANSPEC data on the basis of classification scheme given by \citet{2004ApJ...608..797O}. A total of 13 Class $\textsc{II}$ YSOs were identified in region R1 whereas 22 YSOs were identified in region R2 using TANSPEC, out of which 9 were Class $\textsc{I}$ and 13 were Class $\textsc{II}$.

In such a manner, we have identified 41 YSOs in the $18^\prime.5 \times18^\prime.5$ region.

\vspace{-2em}




{
\footnotesize
\bibliography{thebibliography}

\begin{thebibliography}{}
\expandafter\ifx\csname natexlab\endcsname\relax\def\natexlab#1{#1}\fi

\bibitem[{{Andr{\'e}} {$et~al$.}(2014){Andr{\'e}}, {Di Francesco},
  {Ward-Thompson}, {Inutsuka}, {Pudritz}, \& {Pineda}}]{2014prpl.conf...27A}
{Andr{\'e}}, P., {Di Francesco}, J., {Ward-Thompson}, D., {$et~al$.} 2014, in
  Protostars and Planets VI, ed. H.~{Beuther}, R.~S. {Klessen}, C.~P.
  {Dullemond}, \& T.~{Henning}, 27

\bibitem[{{Andr{\'e}} {$et~al$.}(2010){Andr{\'e}}, {Men'shchikov}, {Bontemps},
  {K{\"o}nyves}, {Motte}, {Schneider}, {Didelon}, {Minier}, {Saraceno},
  {Ward-Thompson}, {di Francesco}, {White}, {Molinari}, {Testi}, {Abergel},
  {Griffin}, {Henning}, {Royer}, {Mer{\'\i}n}, {Vavrek}, {Attard},
  {Arzoumanian}, {Wilson}, {Ade}, {Aussel}, {Baluteau}, {Benedettini},
  {Bernard}, {Blommaert}, {Cambr{\'e}sy}, {Cox}, {di Giorgio}, {Hargrave},
  {Hennemann}, {Huang}, {Kirk}, {Krause}, {Launhardt}, {Leeks}, {Le Pennec},
  {Li}, {Martin}, {Maury}, {Olofsson}, {Omont}, {Peretto}, {Pezzuto}, {Prusti},
  {Roussel}, {Russeil}, {Sauvage}, {Sibthorpe}, {Sicilia-Aguilar}, {Spinoglio},
  {Waelkens}, {Woodcraft}, \& {Zavagno}}]{2010A&A...518L.102A}
{Andr{\'e}}, P., {Men'shchikov}, A., {Bontemps}, S., {$et~al$.} 2010, \aap,
  518, L102

\bibitem[{{Bastien}(1983)}]{1983A&A...119..109B}
{Bastien}, P. 1983, \aap, 119, 109

\bibitem[{Bessell \& Brett(1988)}]{Bessell_1988}
Bessell, M.~S., \& Brett, J.~M. 1988, Publications of the Astronomical Society
  of the Pacific, 100, 1134

\bibitem[{{Bhadari} {$et~al$.}(2022){Bhadari}, {Dewangan}, {Ojha}, {Pirogov},
  \& {Maity}}]{2022ApJ...930..169B}
{Bhadari}, N.~K., {Dewangan}, L.~K., {Ojha}, D.~K., {Pirogov}, L.~E., \&
  {Maity}, A.~K. 2022, \apj, 930, 169

\bibitem[{Bhadari {$et~al$.}(2020)Bhadari, Dewangan, Pirogov, \&
  Ojha}]{Bhadari_2020}
Bhadari, N.~K., Dewangan, L.~K., Pirogov, L.~E., \& Ojha, D.~K. 2020, The
  Astrophysical Journal, 899, 167

\bibitem[{{Bossini} {$et~al$.}(2019){Bossini}, {Vallenari}, {Bragaglia},
  {Cantat-Gaudin}, {Sordo}, {Balaguer-N{\'u}{\~n}ez}, {Jordi}, {Moitinho},
  {Soubiran}, {Casamiquela}, {Carrera}, \& {Heiter}}]{2019A&A...623A.108B}
{Bossini}, D., {Vallenari}, A., {Bragaglia}, A., {$et~al$.} 2019, \aap, 623,
  A108

\bibitem[{{Cambr{\'e}sy}(1999)}]{1999A&A...345..965C}
{Cambr{\'e}sy}, L. 1999, \aap, 345, 965

\bibitem[{{Cambr{\'e}sy} {$et~al$.}(2002){Cambr{\'e}sy}, {Beichman}, {Jarrett},
  \& {Cutri}}]{2002AJ....123.2559C}
{Cambr{\'e}sy}, L., {Beichman}, C.~A., {Jarrett}, T.~H., \& {Cutri}, R.~M.
  2002, \aj, 123, 2559

\bibitem[{Churchwell {$et~al$.}(2006)Churchwell, Povich, Allen, Taylor, Meade,
  Babler, Indebetouw, Watson, Whitney, Wolfire, Bania, Benjamin, Clemens,
  Cohen, Cyganowski, Jackson, Kobulnicky, Mathis, Mercer, Stolovy, Uzpen,
  Watson, \& Wolff}]{Churchwell_2006}
Churchwell, E., Povich, M.~S., Allen, D., {$et~al$.} 2006, The Astrophysical
  Journal, 649, 759

\bibitem[{Churchwell {$et~al$.}(2007)Churchwell, Watson, Povich, Taylor,
  Babler, Meade, Benjamin, Indebetouw, \& Whitney}]{Churchwell_2007}
Churchwell, E., Watson, D.~F., Povich, M.~S., {$et~al$.} 2007, The
  Astrophysical Journal, 670, 428

\bibitem[{{Condon} {$et~al$.}(1998){Condon}, {Cotton}, {Greisen}, {Yin},
  {Perley}, {Taylor}, \& {Broderick}}]{1998AJ....115.1693C}
{Condon}, J.~J., {Cotton}, W.~D., {Greisen}, E.~W., {$et~al$.} 1998, \aj, 115,
  1693

\bibitem[{{Deharveng} {$et~al$.}(2010){Deharveng}, {Schuller}, {Anderson},
  {Zavagno}, {Wyrowski}, {Menten}, {Bronfman}, {Testi}, {Walmsley}, \&
  {Wienen}}]{2010A&A...523A...6D}
{Deharveng}, L., {Schuller}, F., {Anderson}, L.~D., {$et~al$.} 2010, \aap, 523,
  A6

\bibitem[{{Dewangan}(2019)}]{2019ApJ...884...84D}
{Dewangan}, L.~K. 2019, \apj, 884, 84

\bibitem[{Dewangan {$et~al$.}(2017)Dewangan, Baug, Ojha, Janardhan, Devaraj, \&
  Luna}]{Dewangan_2017}
Dewangan, L.~K., Baug, T., Ojha, D.~K., {$et~al$.} 2017, The Astrophysical
  Journal, 845, 34

\bibitem[{{Dewangan} {$et~al$.}(2022{\natexlab{a}}){Dewangan}, {Bhadari},
  {Maity}, {Pandey}, {Sharma}, {Baug}, \& {Eswaraiah}}]{2022arXiv221004658D}
{Dewangan}, L.~K., {Bhadari}, N.~K., {Maity}, A.~K., {$et~al$.}
  2022{\natexlab{a}}, arXiv e-prints, arXiv:2210.04658

\bibitem[{{Dewangan} {$et~al$.}(2021){Dewangan}, {Dhanya}, {Bhadari}, {Ojha},
  \& {Baug}}]{2021MNRAS.506.6081D}
{Dewangan}, L.~K., {Dhanya}, J.~S., {Bhadari}, N.~K., {Ojha}, D.~K., \& {Baug},
  T. 2021, \mnras, 506, 6081

\bibitem[{{Dewangan} {$et~al$.}(2016){Dewangan}, {Ojha}, {Luna}, {Anandarao},
  {Ninan}, {Mallick}, \& {Mayya}}]{2016ApJ...819...66D}
{Dewangan}, L.~K., {Ojha}, D.~K., {Luna}, A., {$et~al$.} 2016, \apj, 819, 66

\bibitem[{Dewangan {$et~al$.}(2020)Dewangan, Ojha, Sharma, del Palacio,
  Bhadari, \& Das}]{Dewangan_2020}
Dewangan, L.~K., Ojha, D.~K., Sharma, S., {$et~al$.} 2020, The Astrophysical
  Journal, 903, 13

\bibitem[{{Dewangan} {$et~al$.}(2017){Dewangan}, {Ojha}, {Zinchenko},
  {Janardhan}, \& {Luna}}]{2017ApJ...834...22D}
{Dewangan}, L.~K., {Ojha}, D.~K., {Zinchenko}, I., {Janardhan}, P., \& {Luna},
  A. 2017, \apj, 834, 22

\bibitem[{Dewangan {$et~al$.}(2019)Dewangan, Pirogov, Ryabukhina, Ojha, \&
  Zinchenko}]{Dewangan_2019}
Dewangan, L.~K., Pirogov, L.~E., Ryabukhina, O.~L., Ojha, D.~K., \& Zinchenko,
  I. 2019, The Astrophysical Journal, 877, 1

\bibitem[{{Dewangan} {$et~al$.}(2022{\natexlab{b}}){Dewangan}, {Zinchenko},
  {Zemlyanukha}, {Liu}, {Su}, {Kurtz}, {Ojha}, {Pazukhin}, \&
  {Mayya}}]{2022ApJ...925...41D}
{Dewangan}, L.~K., {Zinchenko}, I.~I., {Zemlyanukha}, P.~M., {$et~al$.}
  2022{\natexlab{b}}, \apj, 925, 41

\bibitem[{{Dias} {$et~al$.}(2002){Dias}, {Alessi}, {Moitinho}, \&
  {L{\'e}pine}}]{2002A&A...389..871D}
{Dias}, W.~S., {Alessi}, B.~S., {Moitinho}, A., \& {L{\'e}pine}, J.~R.~D. 2002,
  \aap, 389, 871

\bibitem[{Evans {$et~al$.}(2009)Evans, Dunham, Jørgensen, Enoch, Merín, van
  Dishoeck, Alcalá, Myers, Stapelfeldt, Huard, Allen, Harvey, van Kempen,
  Blake, Koerner, Mundy, Padgett, \& Sargent}]{Evans_2009}
Evans, N.~J., Dunham, M.~M., Jørgensen, J.~K., {$et~al$.} 2009, The
  Astrophysical Journal Supplement Series, 181, 321

\bibitem[{{Friel} {$et~al$.}(2014){Friel}, {Donati}, {Bragaglia}, {Jacobson},
  {Magrini}, {Prisinzano}, {Randich}, {Tosi}, {Cantat-Gaudin}, {Vallenari},
  {Smiljanic}, {Carraro}, {Sordo}, {Maiorca}, {Tautvai{\v{s}}ien{\.{e}}},
  {Sestito}, {Zaggia}, {Jim{\'e}nez-Esteban}, {Gilmore}, {Jeffries}, {Alfaro},
  {Bensby}, {Koposov}, {Korn}, {Pancino}, {Recio-Blanco}, {Franciosini},
  {Hill}, {Jackson}, {de Laverny}, {Morbidelli}, {Sacco}, {Worley},
  {Hourihane}, {Costado}, {Jofr{\'e}}, \& {Lind}}]{2014A&A...563A.117F}
{Friel}, E.~D., {Donati}, P., {Bragaglia}, A., {$et~al$.} 2014, \aap, 563, A117

\bibitem[{Goicoechea {$et~al$.}(2015)Goicoechea, Teyssier, Etxaluze, Goldsmith,
  Ossenkopf, Gerin, Bergin, Black, Cernicharo, Cuadrado, Encrenaz, Falgarone,
  Fuente, Hacar, Lis, Marcelino, Melnick, Müller, Persson, Pety, Röllig,
  Schilke, Simon, Snell, \& Stutzki}]{Goicoechea_2015}
Goicoechea, J.~R., Teyssier, D., Etxaluze, M., {$et~al$.} 2015, The
  Astrophysical Journal, 812, 75

\bibitem[{{Golay}(1974)}]{1974ASSL...41.....G}
{Golay}, M., ed. 1974, Astrophysics and Space Science Library, Vol.~41,
  {Introduction to astronomical photometry}

\bibitem[{{Gutermuth} {$et~al$.}(2009){Gutermuth}, {Megeath}, {Myers}, {Allen},
  {Pipher}, \& {Fazio}}]{2009ApJS..184...18G}
{Gutermuth}, R.~A., {Megeath}, S.~T., {Myers}, P.~C., {$et~al$.} 2009, \apjs,
  184, 18

\bibitem[{{Gutermuth} {$et~al$.}(2005){Gutermuth}, {Megeath}, {Pipher},
  {Williams}, {Allen}, {Myers}, \& {Raines}}]{2005ApJ...632..397G}
{Gutermuth}, R.~A., {Megeath}, S.~T., {Pipher}, J.~L., {$et~al$.} 2005, \apj,
  632, 397

\bibitem[{{Henshaw} {$et~al$.}(2013){Henshaw}, {Caselli}, {Fontani},
  {Jim{\'e}nez-Serra}, {Tan}, \& {Hernandez}}]{2013MNRAS.428.3425H}
{Henshaw}, J.~D., {Caselli}, P., {Fontani}, F., {$et~al$.} 2013, \mnras, 428,
  3425

\bibitem[{{Hoemann} {$et~al$.}(2022){Hoemann}, {Heigl}, \&
  {Burkert}}]{2022arXiv220307002H}
{Hoemann}, E., {Heigl}, S., \& {Burkert}, A. 2022, arXiv e-prints,
  arXiv:2203.07002

\bibitem[{{Jose} {$et~al$.}(2013){Jose}, {Pandey}, {Samal}, {Ojha}, {Ogura},
  {Kim}, {Kobayashi}, {Goyal}, {Chauhan}, \& {Eswaraiah}}]{2013MNRAS.432.3445J}
{Jose}, J., {Pandey}, A.~K., {Samal}, M.~R., {$et~al$.} 2013, \mnras, 432, 3445

\bibitem[{{Kaur} {$et~al$.}(2020){Kaur}, {Sharma}, {Dewangan}, {Ojha},
  {Durgapal}, \& {Panwar}}]{2020ApJ...896...29K}
{Kaur}, H., {Sharma}, S., {Dewangan}, L.~K., {$et~al$.} 2020, \apj, 896, 29

\bibitem[{{Kharchenko} {$et~al$.}(2016){Kharchenko}, {Piskunov}, {Schilbach},
  {R{\"o}ser}, \& {Scholz}}]{2016A&A...585A.101K}
{Kharchenko}, N.~V., {Piskunov}, A.~E., {Schilbach}, E., {R{\"o}ser}, S., \&
  {Scholz}, R.~D. 2016, \aap, 585, A101

\bibitem[{{Koenig} \& {Leisawitz}(2014)}]{2014ApJ...791..131K}
{Koenig}, X.~P., \& {Leisawitz}, D.~T. 2014, \apj, 791, 131

\bibitem[{{Kronberger} {$et~al$.}(2006){Kronberger}, {Teutsch}, {Alessi},
  {Steine}, {Ferrero}, {Graczewski}, {Juchert}, {Patchick}, {Riddle},
  {Saloranta}, {Schoenball}, \& {Watson}}]{2006A&A...447..921K}
{Kronberger}, M., {Teutsch}, P., {Alessi}, B., {$et~al$.} 2006, \aap, 447, 921

\bibitem[{{Kumar} {$et~al$.}(2018){Kumar}, {Omar}, {Maheswar}, {Pandey},
  {Sagar}, {Uddin}, {Sanwal}, {Bangia}, {Kumar}, {Yadav}, {Sahu}, {Pant},
  {Reddy}, {Gupta}, {Chand}, {Pandey}, {Joshi}, {Jaiswar}, {Nanjappa},
  {Purushottam}, {Yadav}, {Sharma}, {Pandey}, {Joshi}, {Joshi}, {Lata},
  {Mehdi}, {Misra}, \& {Singh}}]{2018BSRSL..87...29K}
{Kumar}, B., {Omar}, A., {Maheswar}, G., {$et~al$.} 2018, Bulletin de la
  Societe Royale des Sciences de Liege, 87, 29

\bibitem[{{Kumar, M. S. N.} {$et~al$.}(2020){Kumar, M. S. N.}, {Palmeirim, P.},
  {Arzoumanian, D.}, \& {Inutsuka, S. I.}}]{refId0}
{Kumar, M. S. N.}, {Palmeirim, P.}, {Arzoumanian, D.}, \& {Inutsuka, S. I.}
  2020, A\&A, 642, A87

\bibitem[{{Lada} {$et~al$.}(1999){Lada}, {Alves}, \&
  {Lada}}]{1999ApJ...512..250L}
{Lada}, C.~J., {Alves}, J., \& {Lada}, E.~A. 1999, \apj, 512, 250

\bibitem[{{Lada} \& {Lada}(2003)}]{2003ARA&A..41...57L}
{Lada}, C.~J., \& {Lada}, E.~A. 2003, \araa, 41, 57

\bibitem[{Lada {$et~al$.}(2010)Lada, Lombardi, \& Alves}]{Lada_2010}
Lada, C.~J., Lombardi, M., \& Alves, J.~F. 2010, The Astrophysical Journal,
  724, 687

\bibitem[{{Landolt}(1992)}]{1992AJ....104..340L}
{Landolt}, A.~U. 1992, \aj, 104, 340

\bibitem[{{Lombardi} {$et~al$.}(2006){Lombardi}, {Alves}, \&
  {Lada}}]{2006A&A...454..781L}
{Lombardi}, M., {Alves}, J., \& {Lada}, C.~J. 2006, \aap, 454, 781

\bibitem[{{Lumsden} {$et~al$.}(2013){Lumsden}, {Hoare}, {Urquhart},
  {Oudmaijer}, {Davies}, {Mottram}, {Cooper}, \& {Moore}}]{2013yCat..22080011L}
{Lumsden}, S.~L., {Hoare}, M.~G., {Urquhart}, J.~S., {$et~al$.} 2013, VizieR
  Online Data Catalog, J/ApJS/208/11

\bibitem[{{Lundquist} {$et~al$.}(2016){Lundquist}, {Kobulnicky}, {Alexander},
  {Kerton}, \& {Arvidsson}}]{2016yCat..17840111L}
{Lundquist}, M.~J., {Kobulnicky}, H.~A., {Alexander}, M.~J., {Kerton}, C.~R.,
  \& {Arvidsson}, K. 2016, VizieR Online Data Catalog, J/ApJ/784/111

\bibitem[{{Ma} {$et~al$.}(2021){Ma}, {Wang}, {Li}, {Lin}, {Sun}, \&
  {Yang}}]{2021yCat..22540003M}
{Ma}, Y., {Wang}, H., {Li}, C., {$et~al$.} 2021, VizieR Online Data Catalog,
  J/ApJS/254/3

\bibitem[{{Maity} {$et~al$.}(2022){Maity}, {Dewangan}, {Sano}, {Tachihara},
  {Fukui}, \& {Bhadari}}]{2022ApJ...934....2M}
{Maity}, A.~K., {Dewangan}, L.~K., {Sano}, H., {$et~al$.} 2022, \apj, 934, 2

\bibitem[{{Marsh} {$et~al$.}(2015){Marsh}, {Whitworth}, \&
  {Lomax}}]{2015MNRAS.454.4282M}
{Marsh}, K.~A., {Whitworth}, A.~P., \& {Lomax}, O. 2015, \mnras, 454, 4282

\bibitem[{{Marsh} {$et~al$.}(2017){Marsh}, {Whitworth}, {Lomax}, {Ragan},
  {Becciani}, {Cambr{\'e}sy}, {Di Giorgio}, {Eden}, {Elia}, {Kacsuk},
  {Molinari}, {Palmeirim}, {Pezzuto}, {Schneider}, {Sciacca}, \&
  {Vitello}}]{2017MNRAS.471.2730M}
{Marsh}, K.~A., {Whitworth}, A.~P., {Lomax}, O., {$et~al$.} 2017, \mnras, 471,
  2730

\bibitem[{{Meyer} {$et~al$.}(1997){Meyer}, {Calvet}, \&
  {Hillenbrand}}]{1997AJ....114..288M}
{Meyer}, M.~R., {Calvet}, N., \& {Hillenbrand}, L.~A. 1997, \aj, 114, 288

\bibitem[{{Molinari} {$et~al$.}(2010{\natexlab{a}}){Molinari}, {Swinyard},
  {Bally}, {Barlow}, {Bernard}, {Martin}, {Moore}, {Noriega-Crespo}, {Plume},
  {Testi}, {Zavagno}, {Abergel}, {Ali}, {Anderson}, {Andr{\'e}}, {Baluteau},
  {Battersby}, {Beltr{\'a}n}, {Benedettini}, {Billot}, {Blommaert}, {Bontemps},
  {Boulanger}, {Brand}, {Brunt}, {Burton}, {Calzoletti}, {Carey}, {Caselli},
  {Cesaroni}, {Cernicharo}, {Chakrabarti}, {Chrysostomou}, {Cohen},
  {Compiegne}, {de Bernardis}, {de Gasperis}, {di Giorgio}, {Elia}, {Faustini},
  {Flagey}, {Fukui}, {Fuller}, {Ganga}, {Garcia-Lario}, {Glenn}, {Goldsmith},
  {Griffin}, {Hoare}, {Huang}, {Ikhenaode}, {Joblin}, {Joncas}, {Juvela},
  {Kirk}, {Lagache}, {Li}, {Lim}, {Lord}, {Marengo}, {Marshall}, {Masi},
  {Massi}, {Matsuura}, {Minier}, {Miville-Desch{\^e}nes}, {Montier}, {Morgan},
  {Motte}, {Mottram}, {M{\"u}ller}, {Natoli}, {Neves}, {Olmi}, {Paladini},
  {Paradis}, {Parsons}, {Peretto}, {Pestalozzi}, {Pezzuto}, {Piacentini},
  {Piazzo}, {Polychroni}, {Pomar{\`e}s}, {Popescu}, {Reach}, {Ristorcelli},
  {Robitaille}, {Robitaille}, {Rod{\'o}n}, {Roy}, {Royer}, {Russeil},
  {Saraceno}, {Sauvage}, {Schilke}, {Schisano}, {Schneider}, {Schuller},
  {Schulz}, {Sibthorpe}, {Smith}, {Smith}, {Spinoglio}, {Stamatellos},
  {Strafella}, {Stringfellow}, {Sturm}, {Taylor}, {Thompson}, {Traficante},
  {Tuffs}, {Umana}, {Valenziano}, {Vavrek}, {Veneziani}, {Viti}, {Waelkens},
  {Ward-Thompson}, {White}, {Wilcock}, {Wyrowski}, {Yorke}, \&
  {Zhang}}]{2010A&A...518L.100M}
{Molinari}, S., {Swinyard}, B., {Bally}, J., {$et~al$.} 2010{\natexlab{a}},
  \aap, 518, L100

\bibitem[{{Molinari} {$et~al$.}(2010{\natexlab{b}}){Molinari}, {Swinyard},
  {Bally}, {Barlow}, {Bernard}, {Martin}, {Moore}, {Noriega-Crespo}, {Plume},
  {Testi}, {Zavagno}, {Abergel}, {Ali}, {Andr{\'e}}, {Baluteau}, {Benedettini},
  {Bern{\'e}}, {Billot}, {Blommaert}, {Bontemps}, {Boulanger}, {Brand},
  {Brunt}, {Burton}, {Campeggio}, {Carey}, {Caselli}, {Cesaroni}, {Cernicharo},
  {Chakrabarti}, {Chrysostomou}, {Codella}, {Cohen}, {Compiegne}, {Davis}, {de
  Bernardis}, {de Gasperis}, {Di Francesco}, {di Giorgio}, {Elia}, {Faustini},
  {Fischera}, {Fukui}, {Fuller}, {Ganga}, {Garcia-Lario}, {Giard}, {Giardino},
  {Glenn}, {Goldsmith}, {Griffin}, {Hoare}, {Huang}, {Jiang}, {Joblin},
  {Joncas}, {Juvela}, {Kirk}, {Lagache}, {Li}, {Lim}, {Lord}, {Lucas},
  {Maiolo}, {Marengo}, {Marshall}, {Masi}, {Massi}, {Matsuura}, {Meny},
  {Minier}, {Miville-Desch{\^e}nes}, {Montier}, {Motte}, {M{\"u}ller},
  {Natoli}, {Neves}, {Olmi}, {Paladini}, {Paradis}, {Pestalozzi}, {Pezzuto},
  {Piacentini}, {Pomar{\`e}s}, {Popescu}, {Reach}, {Richer}, {Ristorcelli},
  {Roy}, {Royer}, {Russeil}, {Saraceno}, {Sauvage}, {Schilke},
  {Schneider-Bontemps}, {Schuller}, {Schultz}, {Shepherd}, {Sibthorpe},
  {Smith}, {Smith}, {Spinoglio}, {Stamatellos}, {Strafella}, {Stringfellow},
  {Sturm}, {Taylor}, {Thompson}, {Tuffs}, {Umana}, {Valenziano}, {Vavrek},
  {Viti}, {Waelkens}, {Ward-Thompson}, {White}, {Wyrowski}, {Yorke}, \&
  {Zhang}}]{2010PASP..122..314M}
---. 2010{\natexlab{b}}, \pasp, 122, 314

\bibitem[{Myers(2009)}]{Myers_2009}
Myers, P.~C. 2009, The Astrophysical Journal, 700, 1609

\bibitem[{Nakamura {$et~al$.}(2012)Nakamura, Miura, Kitamura, Shimajiri,
  Kawabe, Akashi, Ikeda, Tsukagoshi, Momose, Nishi, \& Li}]{Nakamura_2012}
Nakamura, F., Miura, T., Kitamura, Y., {$et~al$.} 2012, The Astrophysical
  Journal, 746, 25

\bibitem[{Nakamura {$et~al$.}(2014)Nakamura, Sugitani, Tanaka, Nishitani,
  Dobashi, Shimoikura, Shimajiri, Kawabe, Yonekura, Mizuno, Kimura, Tokuda,
  Kozu, Okada, Hasegawa, Ogawa, Kameno, Shinnaga, Momose, Nakajima, Onishi,
  Maezawa, Hirota, Takano, Iono, Kuno, \& Yamamoto}]{Nakamura_2014}
Nakamura, F., Sugitani, K., Tanaka, T., {$et~al$.} 2014, The Astrophysical
  Journal Letters, 791, L23

\bibitem[{{Ojha} {$et~al$.}(2004){Ojha}, {Tamura}, {Nakajima}, {Fukagawa},
  {Sugitani}, {Nagashima}, {Nagayama}, {Nagata}, {Sato}, {Pickles}, \&
  {Ogura}}]{2004ApJ...608..797O}
{Ojha}, D.~K., {Tamura}, M., {Nakajima}, Y., {$et~al$.} 2004, \apj, 608, 797

\bibitem[{{Orlando} {$et~al$.}(2011){Orlando}, {Reale}, {Peres}, \&
  {Mignone}}]{2011MNRAS.415.3380O}
{Orlando}, S., {Reale}, F., {Peres}, G., \& {Mignone}, A. 2011, \mnras, 415,
  3380

\bibitem[{{Pandey} {$et~al$.}(2020){Pandey}, {Sharma}, {Panwar}, {Dewangan},
  {Ojha}, {Bisen}, {Sinha}, {Ghosh}, \& {Pandey}}]{2020ApJ...891...81P}
{Pandey}, R., {Sharma}, S., {Panwar}, N., {$et~al$.} 2020, \apj, 891, 81

\bibitem[{{Panwar} {$et~al$.}(2020){Panwar}, {Sharma}, {Ojha}, {Baug},
  {Dewangan}, {Bhatt}, \& {Pandey}}]{2020ApJ...905...61P}
{Panwar}, N., {Sharma}, S., {Ojha}, D.~K., {$et~al$.} 2020, \apj, 905, 61

\bibitem[{{Pastorelli} {$et~al$.}(2019){Pastorelli}, {Marigo}, {Girardi},
  {Chen}, {Rubele}, {Trabucchi}, {Aringer}, {Bladh}, {Bressan},
  {Montalb{\'a}n}, {Boyer}, {Dalcanton}, {Eriksson}, {Groenewegen},
  {H{\"o}fner}, {Lebzelter}, {Nanni}, {Rosenfield}, {Wood}, \&
  {Cioni}}]{2019MNRAS.485.5666P}
{Pastorelli}, G., {Marigo}, P., {Girardi}, L., {$et~al$.} 2019, \mnras, 485,
  5666

\bibitem[{Pecaut \& Mamajek(2013)}]{Pecaut_2013}
Pecaut, M.~J., \& Mamajek, E.~E. 2013, The Astrophysical Journal Supplement
  Series, 208, 9

\bibitem[{{Perren} {$et~al$.}(2015){Perren}, {V{\'a}zquez}, \&
  {Piatti}}]{2015A&A...576A...6P}
{Perren}, G.~I., {V{\'a}zquez}, R.~A., \& {Piatti}, A.~E. 2015, \aap, 576, A6

\bibitem[{{Phelps} \& {Janes}(1994)}]{1994ApJS...90...31P}
{Phelps}, R.~L., \& {Janes}, K.~A. 1994, \apjs, 90, 31

\bibitem[{Pon {$et~al$.}(2012)Pon, Toalá, Johnstone, Vázquez-Semadeni,
  Heitsch, \& Gómez}]{Pon_2012}
Pon, A., Toalá, J.~A., Johnstone, D., {$et~al$.} 2012, The Astrophysical
  Journal, 756, 145

\bibitem[{{Rieke} \& {Lebofsky}(1985)}]{1985ApJ...288..618R}
{Rieke}, G.~H., \& {Lebofsky}, M.~J. 1985, \apj, 288, 618

\bibitem[{{Sagar} {$et~al$.}(2012){Sagar}, {Kumar}, {Omar}, \&
  {Joshi}}]{2012ASInC...4..173S}
{Sagar}, R., {Kumar}, B., {Omar}, A., \& {Joshi}, Y.~C. 2012, in Astronomical
  Society of India Conference Series, Vol.~4, Astronomical Society of India
  Conference Series, 173

\bibitem[{{Sharma} {$et~al$.}(2008){Sharma}, {Pandey}, {Ogura}, {Aoki},
  {Pandey}, {Sandhu}, \& {Sagar}}]{2008AJ....135.1934S}
{Sharma}, S., {Pandey}, A.~K., {Ogura}, K., {$et~al$.} 2008, \aj, 135, 1934

\bibitem[{Sharma {$et~al$.}(2006)Sharma, Pandey, Ogura, Mito, Tarusawa, \&
  Sagar}]{Sharma_2006}
Sharma, S., Pandey, A.~K., Ogura, K., {$et~al$.} 2006, The Astronomical
  Journal, 132, 1669

\bibitem[{{Sharma} {$et~al$.}(2017){Sharma}, {Pandey}, {Ojha}, {Bhatt},
  {Ogura}, {Kobayashi}, {Yadav}, \& {Pandey}}]{2017MNRAS.467.2943S}
{Sharma}, S., {Pandey}, A.~K., {Ojha}, D.~K., {$et~al$.} 2017, \mnras, 467,
  2943

\bibitem[{{Sharma} {$et~al$.}(2007){Sharma}, {Pandey}, {Ojha}, {Chen}, {Ghosh},
  {Bhatt}, {Maheswar}, \& {Sagar}}]{2007MNRAS.380.1141S}
---. 2007, \mnras, 380, 1141

\bibitem[{Sharma {$et~al$.}(2012)Sharma, Pandey, Pandey, Chauhan, Ogura, Ojha,
  Borrissova, Mito, Verdugo, \& Bhatt}]{10.1093/pasj/64.5.107}
Sharma, S., Pandey, A.~K., Pandey, J.~C., {$et~al$.} 2012, Publications of the
  Astronomical Society of Japan, 64,
  https://academic.oup.com/pasj/article-pdf/64/5/107/17456880/pasj64-0107.pdf,
  107

\bibitem[{{Sharma} {$et~al$.}(2016){Sharma}, {Pandey}, {Borissova}, {Ojha},
  {Ivanov}, {Ogura}, {Kobayashi}, {Kurtev}, {Gopinathan}, \& {Kesh
  Yadav}}]{2016AJ....151..126S}
{Sharma}, S., {Pandey}, A.~K., {Borissova}, J., {$et~al$.} 2016, \aj, 151, 126

\bibitem[{{Sharma} {$et~al$.}(2022){Sharma}, {Ojha}, {Ghosh}, {Ninan}, {Ghosh},
  {Ghosh}, {Manoj}, {Naik}, {D'Costa}, {Krishna Reddy}, {Nanjappa}, {Pandey},
  {Sinha}, {Panwar}, {Antony}, {Kaur}, {Sahu}, {Bangia}, {Poojary}, {Jadhav},
  {Bhagat}, {Meshram}, {Shah}, {Rayner}, {Toomey}, {Sandimani}, \& {Pradeep
  R.}}]{2022PASP..134h5002S}
{Sharma}, S., {Ojha}, D.~K., {Ghosh}, A., {$et~al$.} 2022, \pasp, 134, 085002

\bibitem[{Sharma {$et~al$.}(2022)Sharma, Ojha, Ghosh, Ninan, Ghosh, Ghosh,
  Manoj, Naik, D'Costa, Reddy, Nanjappa, Pandey, Sinha, Panwar, Antony, Kaur,
  Sahu, Bangia, Poojary, Jadhav, Bhagat, Meshram, Shah, Rayner, Toomey, , \&
  Sandimani}]{Sharma_2022}
Sharma, S., Ojha, D.~K., Ghosh, A., {$et~al$.} 2022, Publications of the
  Astronomical Society of the Pacific, 134, 085002

\bibitem[{{Siess} {$et~al$.}(2000){Siess}, {Dufour}, \&
  {Forestini}}]{2000A&A...358..593S}
{Siess}, L., {Dufour}, E., \& {Forestini}, M. 2000, \aap, 358, 593

\bibitem[{{Skrutskie} {$et~al$.}(2006){Skrutskie}, {Cutri}, {Stiening},
  {Weinberg}, {Schneider}, {Carpenter}, {Beichman}, {Capps}, {Chester},
  {Elias}, {Huchra}, {Liebert}, {Lonsdale}, {Monet}, {Price}, {Seitzer},
  {Jarrett}, {Kirkpatrick}, {Gizis}, {Howard}, {Evans}, {Fowler}, {Fullmer},
  {Hurt}, {Light}, {Kopan}, {Marsh}, {McCallon}, {Tam}, {Van Dyk}, \&
  {Wheelock}}]{2006AJ....131.1163S}
{Skrutskie}, M.~F., {Cutri}, R.~M., {Stiening}, R., {$et~al$.} 2006, \aj, 131,
  1163

\bibitem[{{Stetson}(1992)}]{1992ASPC...25..297S}
{Stetson}, P.~B. 1992, in Astronomical Society of the Pacific Conference
  Series, Vol.~25, Astronomical Data Analysis Software and Systems I, ed. D.~M.
  {Worrall}, C.~{Biemesderfer}, \& J.~{Barnes}, 297

\bibitem[{{Tadross}(2009)}]{2009Ap&SS.323..383T}
{Tadross}, A.~L. 2009, \apss, 323, 383

\bibitem[{{Taylor} {$et~al$.}(2003){Taylor}, {Gibson}, {Peracaula}, {Martin},
  {Landecker}, {Brunt}, {Dewdney}, {Dougherty}, {Gray}, {Higgs}, {Kerton},
  {Knee}, {Kothes}, {Purton}, {Uyaniker}, {Wallace}, {Willis}, \&
  {Durand}}]{2003AJ....125.3145T}
{Taylor}, A.~R., {Gibson}, S.~J., {Peracaula}, M., {$et~al$.} 2003, \aj, 125,
  3145

\bibitem[{Wang {$et~al$.}(2020)Wang, Koch, Galván-Madrid, Lai, Liu, Lin, \&
  Pattle}]{Wang_2020}
Wang, J.-W., Koch, P.~M., Galván-Madrid, R., {$et~al$.} 2020, The
  Astrophysical Journal, 905, 158

\bibitem[{{Wang} \& {Hwang}(2020)}]{2020A&A...641A..24W}
{Wang}, T.-M., \& {Hwang}, C.-Y. 2020, \aap, 641, A24

\bibitem[{{Wright} {$et~al$.}(2010){Wright}, {Eisenhardt}, {Mainzer},
  {Ressler}, {Cutri}, {Jarrett}, {Kirkpatrick}, {Padgett}, {McMillan},
  {Skrutskie}, {Stanford}, {Cohen}, {Walker}, {Mather}, {Leisawitz}, {Gautier},
  {McLean}, {Benford}, {Lonsdale}, {Blain}, {Mendez}, {Irace}, {Duval}, {Liu},
  {Royer}, {Heinrichsen}, {Howard}, {Shannon}, {Kendall}, {Walsh}, {Larsen},
  {Cardon}, {Schick}, {Schwalm}, {Abid}, {Fabinsky}, {Naes}, \&
  {Tsai}}]{2010AJ....140.1868W}
{Wright}, E.~L., {Eisenhardt}, P. R.~M., {Mainzer}, A.~K., {$et~al$.} 2010,
  \aj, 140, 1868

\end{thebibliography}
}
\end{document}